\documentclass[seceq]{ptptex}
\usepackage{wrapft}
\usepackage{latexsym}
\usepackage{graphicx}
\usepackage{amsmath} 

\title{
Effects of white noise on parametric resonance in $\lambda \phi^{4}$ theory
}
\author{
Masamichi \textsc{ishihara}%
\footnote{Electronic address: m\_isihar@koriyama-kgc.ac.jp}
}

\inst{
Department of Human Life Studies, Koriyama Women's University \\
Koriyama, 963-8503, Japan 
}
\recdate{\today}

\abst{%
We investigate the effects of white noise on parametric resonance in $\lambda \phi^{4}$ theory.
The potential $V(\phi)$ in this study is 
$\frac{1}{2} m^{2} \phi^{2} + \frac{1}{3} g \phi^{3} + \frac{1}{4} \lambda \phi^{4}$.
An Mathieu-like equation is derived and the derived equation is applied to a partially thermalized system.
The magnitudes of the amplifications are extracted
by solving the equations numerically for various values of parameters.
It is found that the amplification is suppressed by white noise in almost all the cases.
However, in some $g=0$ cases,
the amplification with white noise is slightly stronger than that without white noise.
In the $g=0$ cases, the fields are always amplified. 
The amplification is maximal at $k_{m} \neq 0$ in some $g=0$ cases.
Contrarily, in the $g = {3 \sqrt{2 \lambda} m}/{2}$ cases, 
the fields for some finite modes are suppressed and the amplification is maximal at $k_{m} \sim 0$
when the amplification occurs. 
It is possible to distinguish by these differences
whether the system is on the $g=0$ state or not. 
}
\begin{document}
\maketitle
\section{Introduction}

Scalar field theories are widely used to describe many physical systems. 
One of them is $\lambda \phi^{4}$ theory that is applied to phase transitions \cite{Kleinert}. 
These scalar theories are also used to describe dynamical phenomena such as  
nucleation, spinodal decomposition and  particle production.
It is expected that the results with $\lambda \phi^{4}$ theory give us many indications 
about the above processes.

Mechanism of particle production is an important topic in various branches of physics.
The energy transfer from the (false) vacuum to nonzero modes has been
investigated in terms of particle production. 
An mechanism of particle production is parametric resonance \cite{Landau,Ishihara2004} that
the periodic oscillation of an field amplifies other fields.
Another mechanism is spinodal decomposition that 
the roll-down of the vacuum from the top of the potential hill 
with negative mass squared amplifies other fields.
Particles produced by these mechanisms determine the following evolution of the system. 
An effect by produced particles is the heating of the system 
as discussed in the study of early universe.
The similar heating is expected in rebreaking processes of the chiral symmetry 
in high energy heavy ion collisions. 

Parametric resonance in the evolution of the universe 
\cite{Son,Finkel,KaiserUniverse,Traschen,Kofman1994,Boyanovsky1995,Shatanov,Boyanovsky,Baacke1997,Greene,Ramsey,Boyanovsky1997,Kofman,Zlatev,Sornborger,Baacke2000,Berges,Zanchin,Zanchin2,Joras,Bassett} 
and in chiral phase transitions 
\cite{Muller,Kaiser,Hiro-Oka,Dumitru,Ishihara4,Ishihara5,Maedan,Tsue}
has been investigated with scalar field theories.
An example of an oscillator which periodically moves is the condensate of a scalar field.  
The oscillation of the condensate amplifies the other fields, which corresponds to particle production.
If the expansion of the system is not too rapid,
these produced particles will be thermalized by collisions between them.
This thermalization corresponds to the heating of the universe.
In chiral phase transitions, pions (or sigma mesons) will be produced and thermalized.
If the system expands rapidly,
the produced particles may not be thermalized, because the distance between produced particles becomes long. 
Such a case may occur in heavy ion collisions, because the system expands one or three dimensionally \cite{Bjorken}.
Therefore, the momentum spectrum may have a peak that corresponds to 
the momentum of the produced particles due to  parametric resonance \cite{Ishihara4}. 

In a realistic physical phenomenon such as chiral phase transition, 
friction and noise appear because of the existence of an environment in general.
For example, 
the linear $\sigma$ model includes only $\pi$ and $\sigma$ mesons.
However, in a heavy ion collisions, there are other particles of different kinds in practice.
Then these particles constitute an environment. 
Even when the theory describes the system completely, 
some fields can be regarded as an environment if those fields are projected out \cite{Greiner,Biro,Rischke,Yabu}.
An candidate for an environment in an effective theory at low energies is 'hard modes';
Friction and noise appear by integrating the hard modes out
if $\lambda \phi^{4}$ theory is regarded as an effective theory at low energies 
and/or if the attention is paid only to the soft modes. 
Therefore, the effects of friction and noise have been investigated in $\lambda \phi^{4}$ theory, 
because the equation of motion has friction and noise terms effectively.

It has been recognized that noise plays an important role in various fields.
An attractive example is stochastic resonance \cite{Gammaitoni} that signal is enhanced by noise.
Some modes on parametric resonance may be enhanced strongly by noise as found 
in the study of stochastic resonance, though noise disturbs a clear signal in general.
The effects of noise on parametric resonance have been studied 
\cite{Zanchin,Zanchin2,Joras,Bassett2}.
In such studies, it is found that noise amplifies the fields on parametric resonance. 
However, it is expected that the amplification of a field depends on the form of the coupling term to noise.
The coupling form between the amplified field $\phi$ and the noise $\xi$ is $\xi \phi$
in Ref. \citen{Zanchin}. However the coupling form in $\lambda \phi^{4}$ theory is different.

In this paper, we study the effects of white noise on parametric resonance in $\lambda \phi^{4}$ theory.
We use a classical equation of motion as a practical tool,
because the amplification of a classical field can be a signal such as particle production 
in quantum mechanical system.
Friction and noise terms are attached by hand to reveal the role of noise. 
In sec. \ref{sec:formulation}, the basic equation that includes the effects of noise is derived.
In subsec. \ref{subsec:eq_of_motion}, 
the equation of motion for the soft modes is derived in the case that  
the condensate moves periodically near the bottom of the potential.
In subsec. \ref{subsec:application}, 
the adopted method in subsec. \ref{subsec:eq_of_motion} is applied to 
the $\lambda \phi^{4}$ theory at finite temperature. 
The derived equation is used to reveal the effects of white noise in the subsequent section. 
In Sec. \ref{sec:numerical}, 
the numerical results are presented for various values of parameters by solving the equation derived in 
Sec. \ref{subsec:application} on the symmetric vacuum and the asymmetric vacuum.
Sec. \ref{sec:conclusions} is devoted to the conclusions.  

\section{Formulation}
\label{sec:formulation}
\subsection{Equation of motion with white noise}
\label{subsec:eq_of_motion}
The Lagrangian in $\lambda \phi^{4}$ theory is 
\begin{eqnarray}
{\cal L} &=& \frac{1}{2} \partial_{\mu} \phi \partial^{\mu} \phi - 
\frac{\lambda}{4} \left( \phi^{2} - v^{2} \right)^{2},
\label{eqn:potential:phi4}
\end{eqnarray}
where $v^{2}$ can be positive or negative.
However, we do not treat the $v^{2}=0$ case in this paper.
Another treatment is necessary in such a case, because the mass term vanishes.
The minimum of the potential is $\phi=0$ for $v^{2} < 0$.
The potential $V(\phi)$ can be rewritten as 
$
V(\phi) = 
 \frac{1}{2} m^{2} \phi^{2} + \frac{1}{4} \lambda \phi^{4} + \frac{1}{4} \lambda v^{4},
$
where $m^{2} = - \lambda v^{2}$. 
On the contrary, the minima of the potential are $\pm v$ for $v^{2} > 0$.
Here, we rewrite the potential $V(\phi)$ in terms of the new field $\tilde{\phi}$ given by 
$\phi = \tilde{\phi} + v$.
The potential is rewritten as 
$
V(\phi) = 
 \frac{1}{2} m^{2} \tilde{\phi}^{2} + \frac{1}{3} g \tilde{\phi}^{3} + \frac{1}{4} \lambda \tilde{\phi}^{4} ,
$
where $m^{2} = 2 \lambda v^{2}$ and $g=3 \lambda v$. 
Therefore we use the following potential in this paper:
\begin{equation}
V(\phi) = 
\frac{1}{2} m^{2} {\phi}^{2} + \frac{1}{3} g {\phi}^{3} + \frac{1}{4} \lambda {\phi}^{4} .
\end{equation}
For this system, the Euler-Lagrange equation is 
\begin{equation}
\Box \phi + m^{2} \phi + g \phi^{2} + \lambda \phi^{3} = 0 ,
\label{eqn:orig}
\end{equation}
where $\Box \equiv \partial_{t}^{2} - \nabla^{2}$. 
We add friction and noise terms by hand which appear through
the projection to soft modes \cite{Greiner,Biro,Rischke}, 
the Caldeira-Leggett method \cite{Yabu},
the decomposition of the effective action \cite{Morikawa}
and so on.
Therefore, for slowly varing fields (i.e. soft modes), 
we use the following equation corresponding to eq. (\ref{eqn:orig}) as a practical tool:  
\begin{equation}
\Box \phi + \eta(t) \partial_{t} \phi + m^{2} \phi 
 + g \phi^{2} + \lambda \phi^{3} = \xi(t),
\label{eqn:basic}
\end{equation}
where $\eta \partial_{t} \phi$ is the friction term and $\xi$ is the noise term.
We investigate the systems described with eq. (\ref{eqn:basic}) in this paper.

Because we study the amplification of the finite modes, 
we divide the field $\phi$ into two parts as follows:
\begin{equation}
\phi = \Phi + \psi,
\label{eqn:divide}
\end{equation}
where $\Phi$ is the spatial average of $\phi$. 
Substituting eq. (\ref{eqn:divide}) into eq. (\ref{eqn:basic}), we obtain
\begin{align}
& \partial_{t}^{2} \Phi + \eta \partial_{t} \Phi 
   + m^{2} \Phi  + g \Phi^{2} + \lambda \Phi^{3} 
\nonumber \\
& \quad 
   + \Box \psi + \eta \partial_{t} \psi 
   + \left( m^{2} + 2 g \Phi + 3\lambda \Phi^{2} \right) \psi  
   + \left( g + 3 \lambda \Phi \right) \psi^{2} + \lambda \psi^{3} = \xi.
\end{align}

If $\psi$ is sufficiently small at the beginning of the evolution, 
the lowest order equation is 
\begin{equation}
 \partial_{t}^{2} \Phi + \eta \partial_{t} \Phi 
   + m^{2} \Phi  + g \Phi^{2} + \lambda \Phi^{3} = \xi 
.
\label{eqn:lowest}
\end{equation}  
In the region that eq. (\ref{eqn:lowest}) holds, 
the equation for $\psi$ is 
\begin{align}
\Box  \psi + \eta \partial_{t} \psi 
&
   + \left( m^{2} + 2 g \Phi + 3\lambda \Phi^{2} \right) \psi  
   + \left( g + 3 \lambda \Phi \right) \psi^{2} + \lambda \psi^{3} = 0
.
\label{eqn:fluctuation}
\end{align}
We try to solve eq. (\ref{eqn:fluctuation}) with the background field $\Phi$.

The equation of motion for $\Phi$ near the bottom of the potential  with small amplitude 
is approximately obtained by neglecting $\Phi^{2}$ and $\Phi^{3}$ terms in eq. (\ref{eqn:lowest}).  
This equation is 
\begin{equation}
\partial_{t}^{2} \Phi + \eta \partial_{t} \Phi + m^{2} \Phi  = \xi
.
\label{eqn:Phi:smallamp}
\end{equation}
The solution of eq. (\ref{eqn:Phi:smallamp}) is described as $\Phi=\Phi_{\rm H}+\Phi_{\rm IH}$,
where $\Phi_{\rm H}$ is the solution of eq. (\ref{eqn:Phi:smallamp}) with $\xi = 0$ and 
$\Phi_{\rm IH}$ with $\xi \neq 0$.
$\Phi_{\rm H}$ in the $\partial_{t} \eta \sim 0$ case is  
\begin{align}
\Phi_{\rm H}(t) 
\sim 
 \exp \left( -\frac{1}{2} \int_{t_{0}}^{t} ds \eta(s) \right)
	\left[ 
	C_{+} \exp \left(\int_{t_{0}}^{t} ds \ \alpha_{+}(s)\right) 
      + C_{-} \exp \left(\int_{t_{0}}^{t} ds \ \alpha_{-}(s)\right) 
	\right],
\label{eqn:homogeneous}
\end{align}
where $\alpha_{\pm}(t) = \pm \left( \eta^{2}/4 - m^{2} \right)^{1/2}$ and 
$C_{\pm}$ are constants.
$C_{\pm}$ are real for $\eta^{2}/4 - m^{2} > 0$ 
and complex with the relation $C_{+} = C_{-}^{*}$
for $\eta^{2}/4 - m^{2} \le 0$,  because $\Phi_{\rm H}$ is real.

$\Phi_{\rm IH}$ is obtained by harmonic analysis. 
$\Phi_{\rm IH}$ and $\xi$ are expanded to Fourier series:
\begin{subequations}
\begin{eqnarray}
\Phi_{\rm IH} &=& \sum_{n=-\infty}^{\infty} \Phi_{n} e^{i \omega_{n} t}
, \label{eqn:Phi:fourier}
\\
\xi &=& \sum_{n=-\infty}^{\infty} \xi_{n} e^{i \omega_{n} t}
,
\label{eqn:xi:fourier}
\end{eqnarray}
\end{subequations}
where $\omega_{n} = 2 \pi n /\tilde{T}$, $n$ is an integer and $\tilde{T}$ is time interval. 
Substituting eqs. (\ref{eqn:Phi:fourier}) and (\ref{eqn:xi:fourier})
into eq. (\ref{eqn:Phi:smallamp}),
we obtain 
\begin{equation}
\Phi_{n} = \frac{\xi_{n}}{-\omega_{n}^{2} + i \eta \omega_{n} + m^{2}},
\label{eqn:Phi_n_xi_n}
\end{equation}
if $\eta$ is independent of time. 
Thus, the approximate solution is expressed with eqs. (\ref{eqn:homogeneous}) and (\ref{eqn:Phi:fourier}) as follows:
 \begin{equation}
	\Phi \sim \Phi_{\rm H} + \Phi_{\rm IH} .
 \label{eqn:expression_of_Phi}
 \end{equation}
We use this solution to solve eq. (\ref{eqn:fluctuation}). 
$\Phi$ is evidently a random variable, because $\xi$ is included in eq. (\ref{eqn:expression_of_Phi}).
In order to obtain the approximate equation for $\psi$ that includes the effects of white noise, 
we replace random variables in eq. (\ref{eqn:fluctuation}) with statistical averages.  
Here, the statistical average is denoted by $\langle \cdots \rangle$.
For example, $\Phi$ in eq. (\ref{eqn:fluctuation}) is replaced with $\langle \Phi \rangle$.

Hereafter, we assume that $\xi$ is white noise: 
$\xi$ satisfies $\langle \xi(t) \rangle = 0$ and 
the power spectrum $I(\omega;\xi)$ of $\xi$ is a constant $I^{\rm c}$.
The statistical averages $\langle \Phi \rangle$ and  $\langle \Phi^{2} \rangle$ are  
\begin{subequations}
\begin{eqnarray}
\langle \Phi \rangle &=& \Phi_{\rm H}
, \\
\langle \Phi^{2} \rangle &=&
   \left( \Phi_{\rm H}\right)^{2} + \langle \left( \Phi_{\rm IH}\right)^{2} \rangle
.
\end{eqnarray}
\end{subequations}
From eq. (\ref{eqn:Phi_n_xi_n}), 
the relation between the power spectrum $I \left( \omega, \Phi_{\rm IH} \right)$ of $\Phi_{\rm IH}$
and that of $\xi$ is 
\begin{equation}
I \left( \omega, \Phi_{\rm IH} \right) = 
\frac{I \left( \omega, \xi \right)}{\left( \omega^{2} - m^{2} \right)^{2} + \eta^{2} \omega^{2}}
.
\end{equation}
Wiener-Khinchin theorem gives the relation between the power spectrum and the correlation function. 
This theorem holds under the condition that the system is stationary, 
that is, the correlation function depends on only time interval. 
This condition does not always hold in the present case, because $\Phi_{H}$ exists.    
This condition  holds approximately if the motion of the condensate is sufficiently slow.
Moreover, 
we treat the cases that the effects of the initial condition for $\Phi_{\rm IH}$ are weak.
This theorem gives the following relation in such a case 
\begin{equation}
\langle \Phi_{\rm IH}(t_{1}) \Phi_{\rm IH}(t_{2}) \rangle = 
I^{c} \int_{-\infty}^{\infty} d\omega
 \frac{e^{i \omega (t_{1}-t_{2})}}{\left( \omega^{2} - m^{2} \right)^{2} + \eta^{2} \omega^{2}}
.
\label{eqn:corr:t1t2}
\end{equation}
We evaluate eq. (\ref{eqn:corr:t1t2}) for $m^{2} > \eta^{2}/4$.  
The integral in eq. (\ref{eqn:corr:t1t2}) is 
\begin{align}
&
\int_{-\infty}^{\infty}  d\omega
\frac{e^{i \omega (t_{1}-t_{2})}}{\left( \omega^{2} - m^{2} \right)^{2} + \eta^{2} \omega^{2}}
= 
\frac{\pi}{m^{2} \eta} 
\exp \left( -\frac{1}{2} \eta \left| t_{1} - t_{2} \right| \right)
\nonumber \\ & \quad \quad \times 
\Biggl\{
\cos \left[ \left(m^{2} - \eta^{2}/4 \right)^{1/2} \left| t_{1} - t_{2} \right| \right]
+
\frac{\eta
\sin \left[ \left(m^{2} - \eta^{2}/4 \right)^{1/2} \left| t_{1} - t_{2} \right| \right]}
{2 \left(m^{2} - \eta^{2}/4 \right)^{1/2} } 
\Biggr\}
.
\end{align}
Therefore, the statistical average of $\left[ \Phi_{\rm IH} \right]^{2}$ is given by  
\begin{equation}
\langle \left[ \Phi_{\rm IH}(t) \right]^{2} \rangle = 
\frac{\pi I^{c}}{m^{2} \eta}.
\label{eqn:relation_PHIIH_Ic}
\end{equation}
From eq. (\ref{eqn:homogeneous}), 
$\Phi_{H}$ can be rewritten as 
\begin{align}
\Phi_{H} = C \cos([m^{2}-\eta^{2}/4]^{1/2} t + \theta)
\exp \left( -\frac{1}{2} \int_{t_{0}}^{t} ds \ \eta(s) \right)
, 
\end{align}
where $C$ and $\theta$ are constants.
We obtain
\begin{align}
\langle 2 g \Phi + 3 \lambda \Phi^{2} \rangle 
& =
   \frac{3\pi\lambda I^{c}}{m^{2} \eta} 
  + \frac{3}{2} \lambda C^{2} \exp \left( - \int_{t_{0}}^{t} ds \eta(s) \right)
\nonumber \\ & \quad
  + 2 g C \exp \left( - \frac{1}{2} \int_{t_{0}}^{t} ds \eta(s) \right)
    \cos \left[ \left( m^{2} -\eta^{2}/4 \right)^{1/2} t + \theta \right]
\nonumber \\ & \quad
  + 3 \lambda C^{2} \exp \left( - \int_{t_{0}}^{t} ds \eta(s) \right)
    \cos \left[ 2 \left( m^{2} -\eta^{2}/4 \right)^{1/2} t + 2 \theta \right]
.
\label{eqn:aveq_for_phiH}
\end{align}

Next, 
we derive the equation for $\psi$ 
when the amplitude of $\psi$ is small.
Higher terms of $\psi$ are ignored in such a case. 
After Fourier transformation, eq. (\ref{eqn:fluctuation}) becomes 
\begin{equation}
\partial_{t}^{2} \psi + \eta \partial_{t} \psi 
   + \left( \vec{k}^{2} + m^{2} + 2 g \Phi + 3\lambda \Phi^{2} \right) \psi  = 0,
\label{eqn:with_friction}
\end{equation}
where $\vec{k}$ represents the mode.
In order to remove $\partial_{t} \psi$ term, we rewrite eq. (\ref{eqn:with_friction}) in terms of 
the new field $\chi$, which is related to $\psi$ as follows:
\begin{equation}
\psi = \exp \left( - \frac{1}{2}\int_{t_{0}}^{t} ds \eta(s) \right) \chi(t). 
\end{equation}
We obtain the equation in the $\partial_{t} \eta(t) \sim 0$ case:
\begin{equation}
\partial_{t}^{2} \chi + \left(
 \vec{k}^{2} + m^{2} - \frac{\eta^{2}}{4} + 2 g \Phi + 3 \lambda \Phi^{2} 
 \right) \chi = 0 .
\label{eqn:mathieulike}
\end{equation}
Replacing $2 g \Phi + 3 \lambda \Phi^{2}$ with $\langle 2 g \Phi + 3 \lambda \Phi^{2} \rangle$ 
and substituting eq. (\ref{eqn:aveq_for_phiH}) into eq. (\ref{eqn:mathieulike}),  
we obtain 
\begin{subequations}
\begin{align}
&
\partial_{t}^{2} \chi + \left\{ a(t,\vec{k})
   - 2 q_{1}(t) \cos \left[ \left( m^{2} - \eta^{2} / 4 \right)^{1/2} t + \theta \right]
\right. \nonumber \\ &  \quad \left.
   - 2 q_{2}(t) \cos \left[ 2 \left( m^{2} - \eta^{2} / 4 \right)^{1/2} t + 2 \theta \right]
\right\} \chi  = 0 ,
\label{eqn:mathieulike2}
\end{align}
where 
\begin{eqnarray}
a(t,\vec{k}) &=& \vec{k}^{2} + m^{2} - \frac{\eta^{2}}{4} 
            + \frac{3\pi \lambda I^{c}}{m^{2} \eta} 
            + \frac{3}{2} C^{2} \lambda \exp \left(-\int_{t_{0}}^{t} ds \eta(s) \right) 
,
\\
2q_{1}(t) &=& - 2 g C \exp \left(- \frac{1}{2} \int_{t_{0}}^{t} ds \eta(s) \right) 
,
\\
2q_{2}(t) &=& - \frac{3}{2} \lambda  C^{2} \exp \left(- \int_{t_{0}}^{t} ds \eta(s) \right) 
.
\end{eqnarray}
\end{subequations}

\noindent
Rewriting eq. (\ref{eqn:mathieulike2}) in terms of the new variable $z$ given by 
$2z=\left( m^{2} - \eta^{2} / 4 \right)^{1/2} t +\theta$,
we obtain
\begin{align}
\partial_{z}^{2} \chi + \Big\{ \tilde{a}(t,\vec{k}) 
   - 2  \tilde{q}_{1}(t) \cos \left( 2 z \right)
   - 2  \tilde{q}_{2}(t) \cos \left( 4 z \right)
\Big\} \chi 
=0,
\label{eqn:final}
\end{align}
where
$\tilde{a} = (4 a)/( m^{2} - \eta^{2} / 4 )$,
$\tilde{q}_{1} = (4 q_{1})/( m^{2} - \eta^{2} / 4 )$ and 
$\tilde{q}_{2} = (4 q_{2})/( m^{2} - \eta^{2} / 4 )$.
We solve eq. (\ref{eqn:final}) numerically in the following section.
Note that $\tilde{q}_{1}$ is equal to zero in the $g = 0$ cases. 
Amplified modes are determined by $\tilde{a}$ for small $\tilde{q}_{1}$ and small $\tilde{q}_{2}$. 
It is found that 
the amplified modes become soft, because noise increases $\tilde{a}$. 
Therefore, some modes may vanish because of noise.

\subsection{Application to a Partially Thermalized System}
\label{subsec:application}
In this subsection, we treat a partially thermalized system in which the hard modes are thermal.  
This situation may be realized after particle production due to spinodal decomposition in early universe
and in chiral symmetry rebreaking process in high energy nucleus-nucleus collisions.
We apply the method described in the previous subsection.

We use $\lambda \phi^{4}$ theory and 
divide the field $\phi$ into three parts,
the condensate $\Phi$, soft modes $\phi_{\rm s}$ and hard modes $\phi_{\rm h}$
\cite{Ishihara5,Mocsy}.
It is assumed that the hard modes are thermal. 
The thermal average of $\phi_{\rm h}^{2n+1}$ vanishes in the free particle approximation
\begin{equation}
\langle \phi_{\rm h}^{2n+1} \rangle_{\rm T} = 0,
\label{eqn:themal_av}
\end{equation}
where $\langle \cdots \rangle_{\rm T}$ indicates the thermal average to the  hard modes
and $n$ is a non-negative integer.
The thermal average of $\lambda \phi^{4}$ potential to the hard modes with eq. (\ref{eqn:themal_av}) is 
\begin{align}
& \langle V(\phi) \rangle_{\rm T} = 
\left\langle \frac{m^{2}}{2} \phi^{2} + \frac{g}{3} \phi^{3} + \frac{\lambda}{4} \phi^{4} \right\rangle_{T} 
\nonumber \\ & \quad 
= g \langle \phi_{\rm h}^{2} \rangle_{\rm T} \Phi + \frac{1}{2} m^{2}(T) \Phi^{2} + \frac{g}{3} \Phi^{3} + \frac{\lambda}{4} \Phi^{4}
+ g \langle \phi_{\rm h}^{2} \rangle_{\rm T} \phi_{\rm s} +\frac{1}{2} m^{2}(T) \phi_{\rm s}^{2} + \frac{g}{3} \phi_{\rm s}^{3} + \frac{\lambda}{4} \phi_{\rm s}^{4} 
\nonumber \\ & \quad  
+ m^{2} \phi_{\rm s} \Phi + g \left( \phi_{\rm s} \Phi^{2} + \phi_{\rm s}^{2} \Phi \right)  
+ \frac{\lambda}{4} \biggl( 
     4 \Phi^{3} \phi_{\rm s} + 6 \Phi^{2} \phi_{\rm s}^{2} 
    + 4 \Phi \phi_{\rm s}^{3}
    + 12 \Phi \phi_{\rm s} \langle \phi_{\rm h}^{2} \rangle_{\rm T} + \langle \phi_{\rm h}^{4} \rangle_{\rm T}  
                    \biggr)
,
\label{eqn:effective_potential}
\end{align}
where $m^{2}(T) = m^{2} + 3 \lambda \langle \phi_{\rm h}^{2} \rangle_{\rm T}$.
Kinetic part is expanded in the same way:
\begin{align}
\frac{1}{2} \left\langle \partial_{\mu} \phi \partial^{\mu} \phi \right\rangle_{\rm T}
= & 
\frac{1}{2} \partial_{\mu} \Phi \partial^{\mu} \Phi 
+ \frac{1}{2} \partial_{\mu} \phi_{\rm s} \partial^{\mu} \phi_{\rm s}  
+ \partial_{\mu} \phi_{\rm s} \partial^{\mu} \Phi
+ \frac{1}{2} \langle  \partial_{\mu} \phi_{\rm h} \partial^{\mu} \phi_{\rm h} \rangle_{\rm T}.
\end{align}
The $\langle \phi_{h}^{2n} \rangle_{T}$ and 
$\langle  \partial_{\mu} \phi_{\rm h} \partial^{\mu} \phi_{\rm h} \rangle_{\rm T}$ terms 
are independent of $\Phi$ and $\phi_{\rm s}$ in the massless free particle approximation 
\cite{Gavin}.
The Euler-Lagrange equation for $\Phi$ and $\phi_{\rm s}$ 
after taking the thermal average to the hard modes is 
\begin{align}
& 
\ddot{\Phi} + m^{2}(T) \Phi + g \Phi^{2} + \lambda \Phi^{3} 
+ g \langle \phi_{\rm h}^{2} \rangle_{\rm T} 
\nonumber \\ & \quad 
+ \Box \phi_{\rm s} + m^{2}(T) \phi_{\rm s}  
+ g \phi_{\rm s}^{2} + \lambda \phi_{\rm s}^{3} 
+ 2 g \phi_{\rm s} \Phi + 3 \lambda \left( \Phi \phi_{\rm s}^{2} + \Phi^{2} \phi_{\rm s} \right) = 0 
.
\end{align}
Here, the friction and noise terms are added by hand as in the previous subsection.
The approximate equations for $\Phi$ and $\phi_{\rm s}$ are, respectively, 
\begin{subequations}
\begin{align}
& \ddot{\Phi} + \eta \dot{\Phi} + m^{2}(T) \Phi + g \Phi^{2} + \lambda \Phi^{3} + g \langle \phi_{\rm h}^{2} \rangle_{\rm T} = \xi , 
\label{eqn:thermal:Phi}
\\
& \Box \phi_{\rm s} + \eta \dot{\phi}_{\rm s} + m^{2}(T) \phi_{\rm s} 
+ ( 2 g \Phi  + 3 \lambda \Phi^{2} ) \phi_{\rm s}  = 0 .
\end{align}
\end{subequations}
In order to remove the $g \langle \phi_{\rm h}^{2} \rangle_{\rm T}$ term in  eq. (\ref{eqn:thermal:Phi}), 
we rewrite eq. (\ref{eqn:thermal:Phi}) in terms of the new field $\Xi$ given by $\Phi = \Xi + \Phi_{0}$, 
%
%
where $\Phi_{0}$ is the solution of the equation 
$\lambda \Phi_{0}^{3} + g \Phi_{0}^{2} + m^{2}(T) \Phi_{0} + g \langle \phi_{\rm h}^{2} \rangle_{\rm T} = 0$. 
According to the calculation described in the previous section, 
we obtain the equation for $\phi_{\rm s}$ by Fourier transformation and changing of the variable:
\begin{subequations}
\begin{equation}
\partial_{z}^{2} {\chi} + \left\{ a^{T}  - 2q_{1}^{T} \cos(2z) - 2q_{2}^{T} \cos(4z) \right\} \chi = 0
, 
\label{eqn:finiteT}
\end{equation}
where 
\begin{align}
 \phi_{s}(t,\vec{k}) &= \exp \left(-\frac{1}{2} \int_{t_{0}}^{t} ds \eta(s) \right) \chi(t,\vec{k}) 
, \\ 
 a^{T} &= 
	4 \Biggl[ \vec{k}^{2} + M^{2}(T) - \frac{\eta^{2}}{4} 
      + \frac{3\pi \lambda I^{c}}{M^{2}(T) \eta} 
      + \frac{3}{2} C^{2} \lambda \exp \left(-\int_{t_{0}}^{t} ds \ \eta(s) \right) \Biggr]
\nonumber \\ & \quad \times
	\Biggl[ M^{2}(T) - \eta^{2} / 4 \Biggr]^{-1}
, \\
 q_{1}^{T} &=  
	- 4 \left( g + 3 \lambda \Phi_{0} \right) C \exp \left(- \frac{1}{2} \int_{t_{0}}^{t} ds \ \eta(s) \right)
	  \left[ M^{2}(T) - \eta^{2} / 4  \right]^{-1}
, \\
 q_{2}^{T} &=  
	- 3 \lambda  C^{2} \exp \left(- \int_{t_{0}}^{t} ds \ \eta(s) \right)
	  \left[ M^{2}(T) - \eta^{2} / 4  \right]^{-1}
, \\
 2z &= \left( M^{2}(T) - \frac{\eta^{2}}{4} \right)^{\frac{1}{2}} t + \theta 
, \\
 M^{2}(T) &= m^{2}(T) + 2 g \Phi_{0} + 3 \lambda \Phi_{0}^{2}
.
\end{align}
\end{subequations}
The remaining task in this subsection is to evaluate $m^{2}(T)$ and $I^{c}$.
$m^{2}(T)$ includes the effects of the hard modes.
$\langle \phi_{\rm h}^{2} \rangle_{\rm T}$ is calculable with the cutoff $\Lambda$ 
between soft and hard modes.
\begin{equation}
m^{2}(T) = m^{2} + \frac{3}{2\pi^{2}} \lambda T^{2} \int_{\Lambda/T}^{\infty} du \frac{u}{e^{u} - 1},
\label{eqn:m2T}
\end{equation}
where the vacuum contribution to $\langle \phi_{\rm h}^{2} \rangle_{\rm T}$ is dropped.
If $\Lambda$ is sufficiently small and/or $T$ is sufficiently  high, 
eq. (\ref{eqn:m2T}) is approximately
\begin{subequations}
\begin{equation}
m^{2}(T) \sim m^{2} + \frac{\lambda}{4} T^{2} .
\label{eqn:m2_at_highT}
\end{equation}
On the contrary, if $\Lambda/T$ is sufficiently large,
\begin{equation}
m^{2}(T) \sim m^{2} + \frac{3 \lambda}{2\pi^{2}} T \Lambda \exp \left( -\Lambda/T \right) .
\end{equation}
\end{subequations}

$I^{c}$ is related to $\langle \Xi_{\rm IH}^{2} \rangle_{T} $ with eq. (\ref{eqn:relation_PHIIH_Ic}). 
We treat the $\Xi_{\rm H} = 0$ case to determine the expression of $I^{c}$.
The energy $E({\Xi})$ of $\Xi$  for small $\phi_{s}$ is approximately 
\begin{equation}
E({\Xi}) = V \left[ \frac{1}{2} \left( \partial_{t} \Xi \right)^{2} + \frac{1}{2} M^{2}(T) \Xi^{2} \right],
\end{equation}
where $V$ is the volume of the system. 
The statistical weight $dw$ is 
\begin{equation}
dw = \frac{M^{2}(T) V}{2\pi T}  \exp\left( - E({\Xi_{\rm IH}})/T \right) d(\partial_{t} \Xi_{\rm IH}) d\Xi_{\rm IH}.
\end{equation}
Thus, the statistical average $\langle \left( \Xi_{\rm IH} \right)^{2} \rangle$ is 
\begin{equation}
\langle \left( \Xi_{\rm IH} \right)^{2} \rangle 
= \int dw \left( \Xi_{\rm IH} \right)^{2}
= \frac{T}{M^{2}(T) V} . 
\end{equation}
With the help of eq. (\ref{eqn:relation_PHIIH_Ic}), we obtain 
\begin{equation}
I^{c} = \frac{\eta T}{\pi V}. 
\label{eqn:expression_of_Ic}
\end{equation}
As shown in Ref. \citen{Biro},
this indicates that the correlation of $\xi$ is represented as follows:
\begin{equation}
\langle \xi(t_{1}) \xi(t_{2}) \rangle = 2 \pi I^{c} \delta (t_{1} - t_{2})
 = \frac{2 \eta T}{V}  \delta (t_{1} - t_{2}).
\label{eqn:randomforce_corr}
\end{equation}
Therefore, the contribution from $\Xi_{\rm IH}$ in $a^{T}$ is $\eta$ independent:
\begin{equation}
\frac{3 \pi \lambda I^{c}}{\left[M(T)\right]^{2} \eta} = \frac{3 \lambda T}{\left[M(T)\right]^{2} V}.
\label{eqn:Icinsert}
\end{equation}
At high $T$, eq. (\ref{eqn:Icinsert}) is proportional to $(VT)^{-1}$ and independent of $\lambda$.
On the other hand, $q_{1}^{T}$ and $q_{2}^{T}$ go to zero as $T$ goes to infinity.
Therefore, the fields are not amplified at extremely high $T$.
In a partially thermalized system with white noise,
eq. (\ref{eqn:finiteT}) with eq. (\ref{eqn:expression_of_Ic}) will be  solved.


\section{Numerical Study}
\label{sec:numerical}
There are many applications of $\lambda \phi^{4}$ theory.
Inflation and chiral phase transitions are interesting examples.
In this section, we study the effects of white noise on parametric resonance generally 
in $\lambda \phi^{4}$ theory by using scaled parameters. 
Then, we can obtain some indications about above subjects from the results if we want. 
We use the parameters scaled by mass $m$, for example, $\vec{k}_{m} = \vec{k}/m$.
The parameters and the variables in the calculations presented below are summarized:
$\chi_{m} = \chi/m $, $T_{m}=T/m$, $\eta_{m} = \eta/m$, $g_{m} = g/m$, $M_{m}(T) = M(T)/m$ and $C_{m} = C / m$.
Though 
$\eta$ or $I_{c}$ cannot be given without clarifying the origin of the dissipation,
 $I_{c}$ is related to $\eta$ through the fluctuation-dissipation theorem. 
Eq. (\ref{eqn:finiteT}) is rewritten with these parameters for $\Lambda/T \sim 0$:
\begin{subequations}
\begin{equation}
\partial_{z}^{2} {\chi_{m}} + \left\{ a^{T}  - 2q_{1}^{T} \cos(2z) - 2q_{2}^{T} \cos(4z) \right\} \chi_{m} = 0, 
\label{eq:numerical}
\end{equation}
where
\begin{align}
 a^{T} &= 4 + 
  4 \Biggl[ \vec{k}_{m}^{2} + \frac{3 \lambda T_{m}}{M_{m}^{2}(T) (m^{3} V)} 
  + \frac{3}{2} \lambda C_{m}^{2} \exp \left( - \int_{w_{0}}^{w} dw \eta_{m}(w) \right) \Biggr]
\notag \\ & \quad\quad 
\times \Biggl[ \left( M_{m}(T)\right)^{2} - \eta_{m}^{2}/4 \Biggr]^{-1}
,
\label{eqn:numerical:aT}
\\
q_{1}^{T} &= -4 \Biggl[g_{m}+ 3 \lambda \left( \frac{\Phi_{0}}{m} \right) \Biggr] C_{m} 
	\exp \left( - \frac{1}{2} \int_{w_{0}}^{w} dw \eta_{m}(w) \right)
\notag \\ & \quad\quad \times
	\Biggl[ \left( M_{m}(T)\right)^{2} - \eta_{m}^{2}/4 \Biggr]^{-1}
,
\label{eqn:q1T}
\\
 q_{2}^{T} &=  3 \lambda  C_{m}^{2} \exp \left(- \int_{w_{0}}^{w} dw \ \eta_{m} \right)
	\left[ \left( M_{m}(T)\right)^{2} - \eta_{m}^{2}/4 \right]^{-1}
,\\
 2z &= \left( M_{m}^{2}(T) - \eta_{m}^{2}/4 \right)^{1/2} w + \theta 
,\\
 \left( M_{m}(T) \right)^{2} &= 1 + \frac{1}{4} \lambda T_{m}^{2} 
 + 2 g_{m} \left( \frac{\Phi_{0}}{m} \right) + 3 \lambda \left( \frac{\Phi_{0}}{m} \right)^{2}
.
\end{align}
\end{subequations}
Here, $w_{0}$ in eq. (\ref{eqn:numerical:aT}) corresponds to $z_{0}$.  
It is expected that amplification occurs around $a^{T} = n^{2}$
if the condition $\left| q_{1}^{T} \right| \gg \left| q_{2}^{T} \right|$ is satisfied,
 where $n$ is a positive integer.
Note that $a^{T}$ is equal to or larger than 4 for positive $\lambda$ if 
$\left( M_{m}(T) \right)^{2} - \eta_{m}^{2}/4 $ is positive.
In this paper, we treat the cases of $\left( M_{m}(T)\right)^{2}  - \eta_{m}^{2}/4 > 0 $.
Eq. (\ref{eq:numerical}) is solved numerically in the calculations presented below.

\begin{figure}[htb]
	\parbox{\halftext}{%
	\includegraphics[width=\halftext]{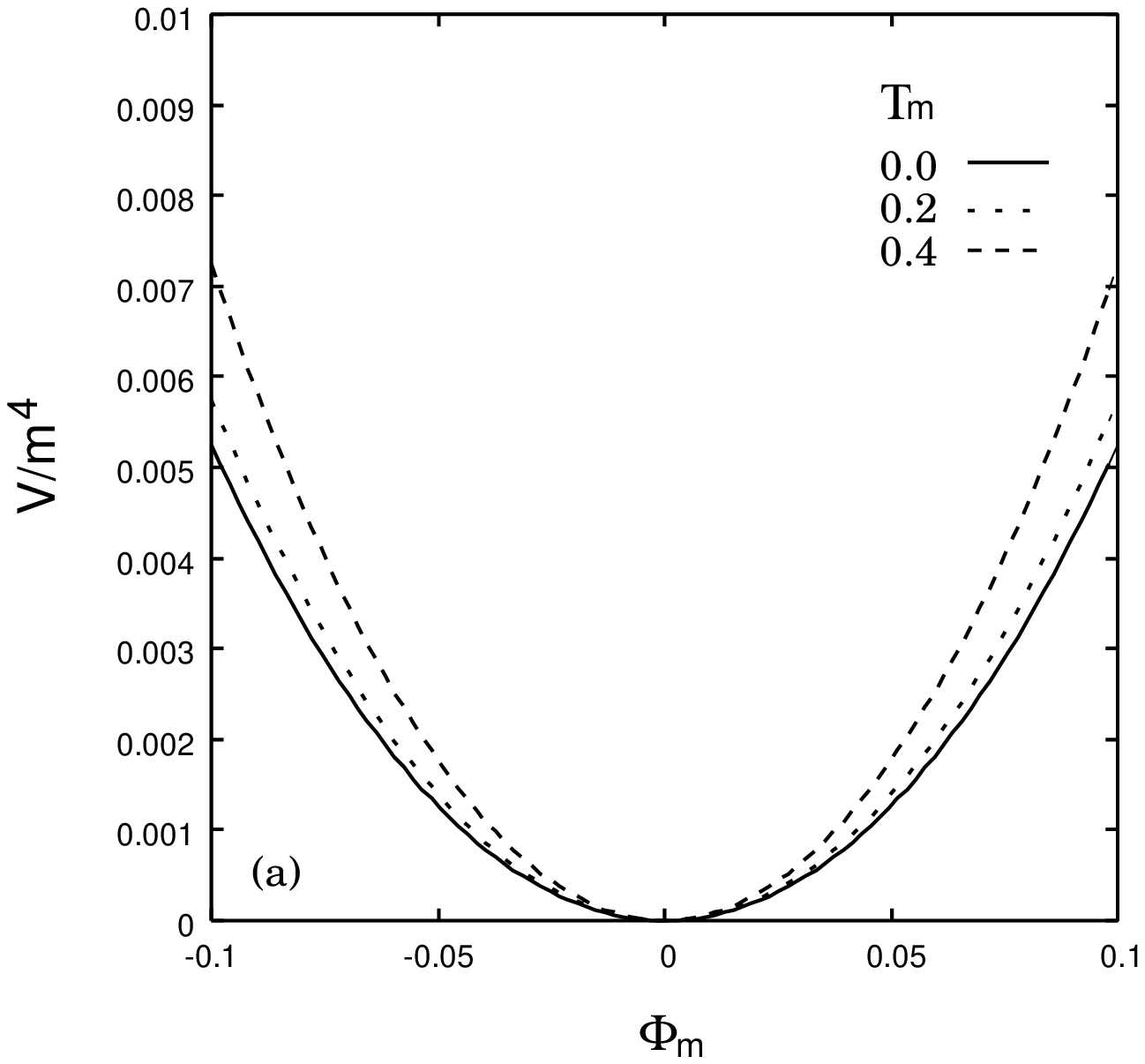}
	}
	\hfill
	\parbox{\halftext}{
	\includegraphics[width=\halftext]{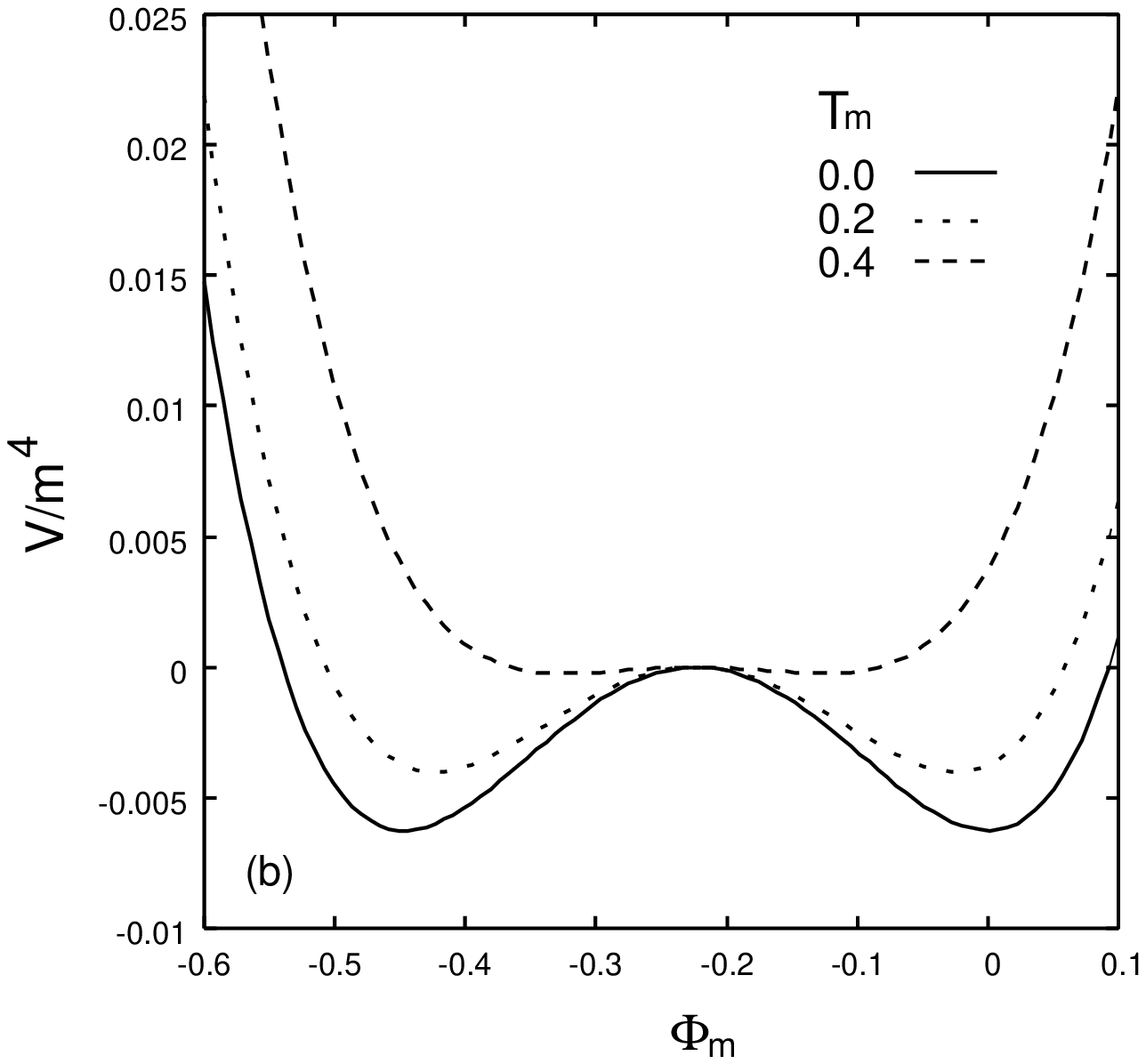}
	}
\caption{
Effective potential scaled by $m^{4}$ for various values of $T_{m}$.  
The values of the parameters are (a) $g_{m}=0, \lambda=10$ and (b) $g_{m} = \frac{3 \sqrt{2 \lambda}}{2}, \lambda=10$.
The energy at $\Phi = -v$ in Fig. (b) is adjusted to zero. 
} 
\label{fig:EP}
\end{figure}
The effective potential is obtained by substituting $\phi_{s}=0$ into eq. (\ref{eqn:effective_potential}).
The effective potential divided by $m^{4}$ 
with eq. (\ref{eqn:m2_at_highT}) is displayed in Fig. \ref{fig:EP}
for (a) $g_{m}=0$, $\lambda=10$ and (b) $g_{m} = \frac{3 \sqrt{2 \lambda}}{2}, \lambda=10$. 
The potential at $\Phi= -v$ is always extremum for $g = 3 \lambda v$ and $m^{2} = 2 \lambda v^{2}$,
which is displayed in Fig. \ref{fig:EP}(b).  
Curves indicate the effective potential for various values of $T_{m}$.
The critical temperature $T_{c}$ that is determined from the effective potential for $v^{2} > 0$ 
is $(2/\lambda)^{1/2} m$.
We discuss the amplification in only the cases of $T \le T_{c}$ for $v^{2} > 0$ in the calculations 
presented below, because the effective potential is convex anywhere for $ T \ge T_{c}$.

In the following subsections, 
we treat the $g_{m} = 0$ and $g_{m} = \frac{3\sqrt{2 \lambda}}{2}$ cases, 
because $g_{m}$ is exactly zero for $v^{2} <  0$  as illustrated in Fig. \ref{fig:EP}(a) 
and 
because $g_{m}$ has a finite value for $v^{2} >  0 $ as in Fig. \ref{fig:EP}(b).
$g_{m}$ is $\frac{3\sqrt{2 \lambda}}{2}$ for the Lagrangian defined by eq. (\ref{eqn:potential:phi4}).
In the numerical calculations of time evolution of the soft modes, 
$\chi_{m}$ is set to 1 at the initial time
and 
the range of the modes $\vec{k}_{m}$ is $0 \le \left| \vec{k}_{m} \right| \le 1.5$.
The condition $\eta \ll m $ is equivalent to $0 \le  \eta_{m} \ll 1$. 
Moreover, we cover two cases: one is that $\eta_{m}$ is independent of $\lambda$ and 
the other is that $\eta_{m}$ is proportional to $\lambda$ \cite{Greiner,Rischke}.
%
We set the initial amplitude $C_{m}$ to $(1-1/\sqrt{3})/\sqrt{2\lambda}$ 
which corresponds to the inflection point of the potential 
at $T_{m}=0$.
This initial amplitude is used in subsec. \ref{subsec:gm=0} and \ref{subsec:gm_ne_0}.
The magnitude of the coupling $\lambda$ is taken to be 
0.1, 1 and 10. 
Then the apparent $\lambda$ dependence may be found.
The remaining parameter is $m^{3} V$ that is related to the magnitude of the noise.
This parameter is included only in the term coming from noise.
In this paper, $m^{3} V$ is set to $10$ if there is no explanation.
$m^{3} V$ is set to $100$  for comparison.
'Amplification' in the figures displayed below represents
the ratio of the amplitude for $z \gg 1$ to that at the initial time.  

\subsection{The $g_{m}=0$ cases}
\label{subsec:gm=0}
In the case of $g_{m}=0$, the $q_{1}^{T}$ term vanishes because of $\Phi_{0}=0$. 
The equation is quite similar to the Mathieu equation,  
except  that $a ^{T}$ and $q_{2}^{T}$ are time dependent functions. 
It is determined from the position of $(a^{T},q_{2}^{T})$ on the $a^{T}$$-$$q_{2}^{T}$ plane
whether the fields are amplified.
To compare eq. (\ref{eq:numerical}) with the Mathieu equation, 
we rewrite eq. (\ref{eq:numerical}) in terms of the new variable $\tilde{z}$ given by $2\tilde{z} = 4z$.
we obtain
\begin{equation}
\partial_{\tilde{z}}^{2} {\chi_{m}} + \left\{ \tilde{a}^{T} - 2 \tilde{q}_{2}^{T} \cos(2\tilde{z}) \right\} \chi_{m} = 0, 
\label{eq:numerical_gm=0}
\end{equation}
where $\tilde{a}^{T} = a^{T}/4$ and $\tilde{q}_{2}^{T} = q_{2}^{T}/4$. 
Large amplification is expected if $\tilde{q}_{2}$ is sufficiently large, because $\tilde{a}^{T}$ can be 1.

\begin{figure}[htb]
	\begin{center}
	\includegraphics[width=\halftext]{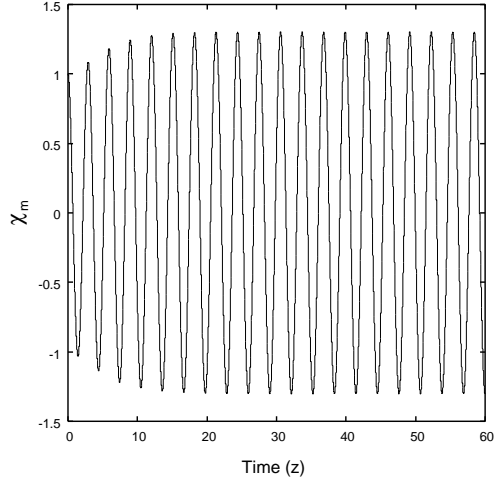}
	\end{center}
\caption{
Time evolution of the soft mode $\chi_{m}$ with $k_{m}=0$.
The values of the parameters are $\lambda=1$, $\eta_{m}=0.1$, $T_{m}=0.1$ and 
$C_{m}= (1-1/\sqrt{3})/\sqrt{2\lambda}$.
} 
\label{fig:time_evolution}
\end{figure}
Typical time evolution of the soft mode $\chi_{m}$ with $k_{m}=0$
is displayed in Fig. \ref{fig:time_evolution} . 
The values of the parameters are $\lambda=1$, $\eta_{m}=0.1$ and $T_{m}=0.1 $. 
The solution for $z \gg 1$ is like a sine function, 
because $q_{2}^{T}$ is almost equal to zero when $z$ is sufficiently large.
Therefore, we can determine the amplitude of $\chi_{m}$ for $z \gg 1$. 
The main purpose of this section is to show the amplification of the soft modes. 
For $z \gg 1$, the amplitudes for various values of the parameters can be determined 
from the numerical solutions of time evolution.

\begin{figure}
	\parbox{\halftext}{%
	\includegraphics[width=\halftext]{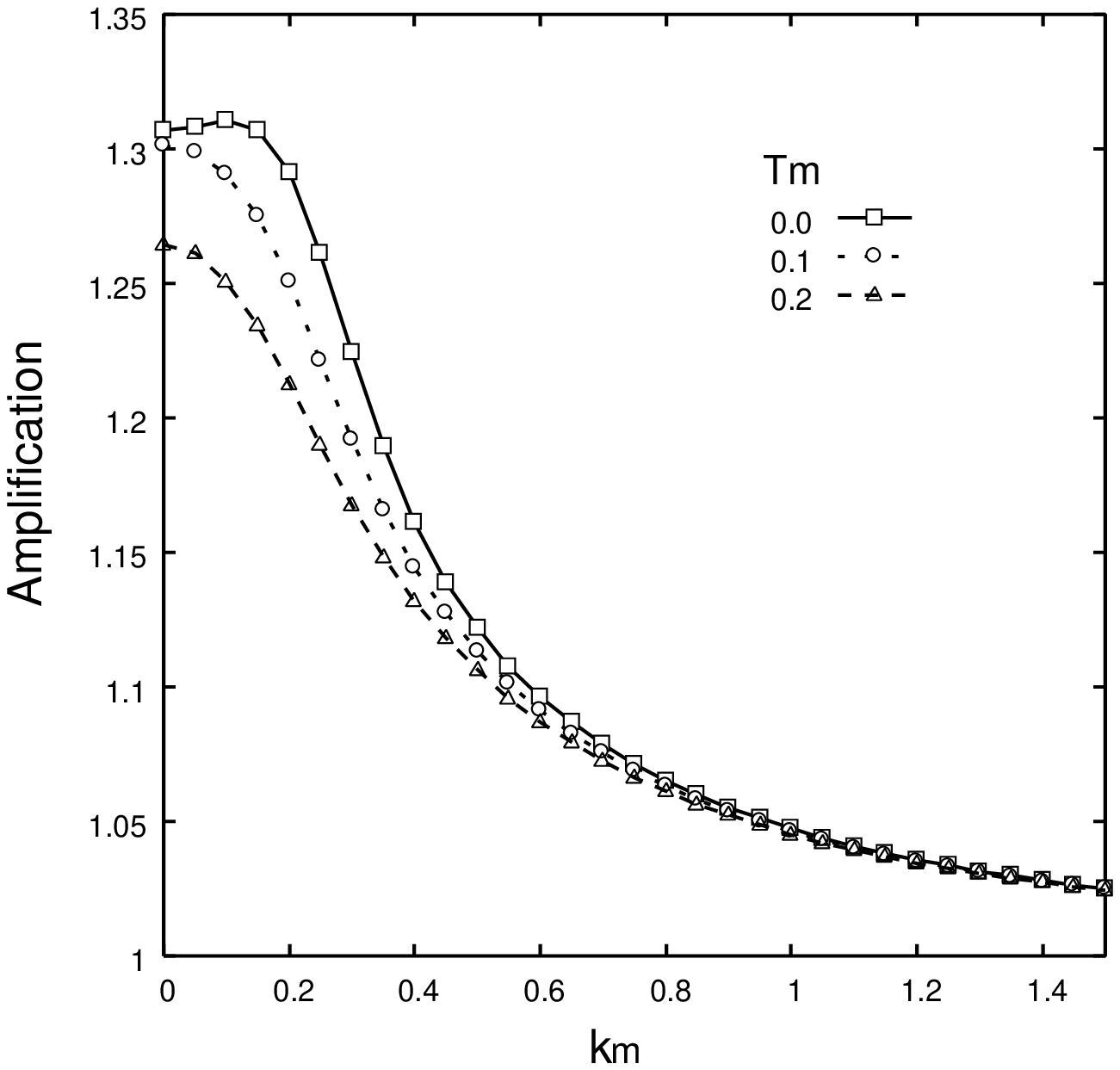}
\caption{Temperature dependence of the amplification with $\lambda=1$ and $\eta_{m} = 0.1$.
The square designates data for $T_{m}=0$, the circle for $T_{m}=0.1$ and the triangle for $T_{m} = 0.2$.
}
\label{fig:g=0:temperature_dep}
	}
	\hfill
	\parbox{\halftext}{%
	\includegraphics[width=\halftext]{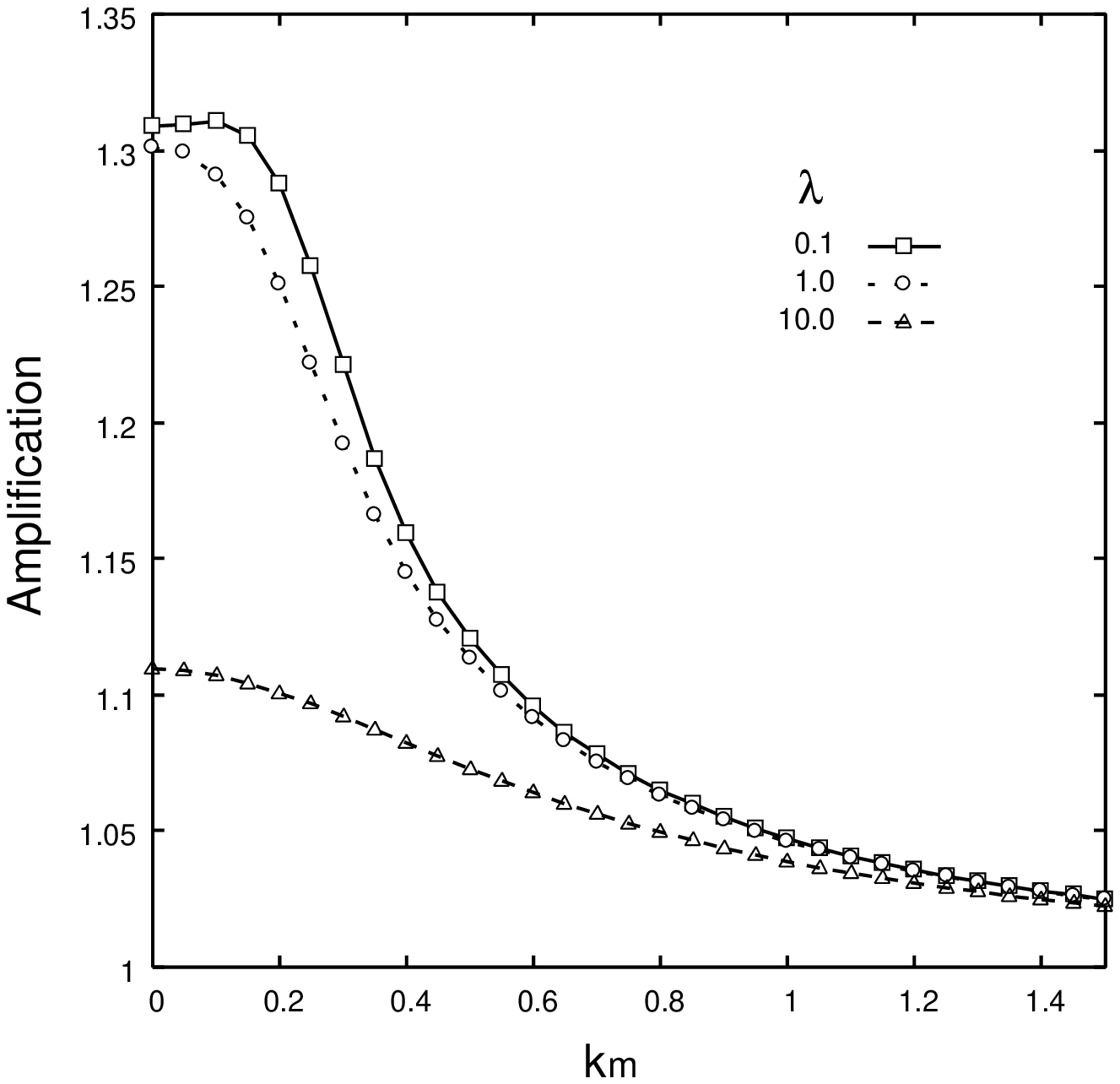}
\caption{Coupling dependence of the amplification with  $\eta_{m} = 0.1$ and $T_{m}=0.1$.
The square designates data for $\lambda=0.1$, the circle for $\lambda=1$ and the triangle for $\lambda=10$.
}
\label{fig:g=0:lambda_dep}
	}
\end{figure}

Figure \ref{fig:g=0:temperature_dep} displays
the temperature dependence of the amplification
with $\lambda=1$ and $\eta_{m}=0.1$.
Each line represents the amplification as a function of $k_{m}$.
It is found from this figure that the amplification is suppressed at high $T_{m}$.
This behavior is explained by the competition between $q_{2}^{T}$ and $a^{T}$.
On one hand, $q_{2}^{T}$ decreases monotonically as $T$ increases;
on the other hand, $a^{T}$ does not vary monotonically as $T$ increases and goes to 4 as $T$ goes to infinity.
Note that the curve of the amplification with white noise at $T_m = 0$ is the same curve 
without white noise at $T_{m}=0$, because the term coming from white noise vanishes at $T_{m}=0$.

The $\lambda$ dependence is displayed in Fig. \ref{fig:g=0:lambda_dep} with $\eta_{m} = 0.1$ and $T_{m}=0.1$. 
It is evident that the fields are amplified for small $\lambda$ and low $k_{m}$. 
This dependence is most likely explained by the $\lambda$ dependence of $a^{T}$ and $q_{2}^{T}$.
$a^{T}$ depends on $\lambda$ explicitly and implicitly.
Because the $\lambda$ dependence of $M_{m}$ at low $T_{m}$ is weak, 
the $\lambda$ dependence of $a^{T}$ is resemble to the function $ c_{1} + c_{2} \lambda$,
where $c_{1}$ and $c_{2}$ are constants. 
(Because $C_{m}$ is $(1 - 1/\sqrt{3})/\sqrt{2\lambda}$ in this calculation,
 the term with $C_{m}$ in $a^{T}$ is independent of $\lambda$.) 
$q^{T}_{2}$ also depends on $\lambda$ weakly through $M_{m}$.
Therefore, the fields are amplified for weak $\lambda$.
Moreover, $a^{T}$ increases monotonically as a function of $k_{m}$. 
Hence, the amplification is strong for small $k_{m}$.

\begin{figure}[htb]
	\parbox{\halftext}{%
	\includegraphics[width=\halftext]{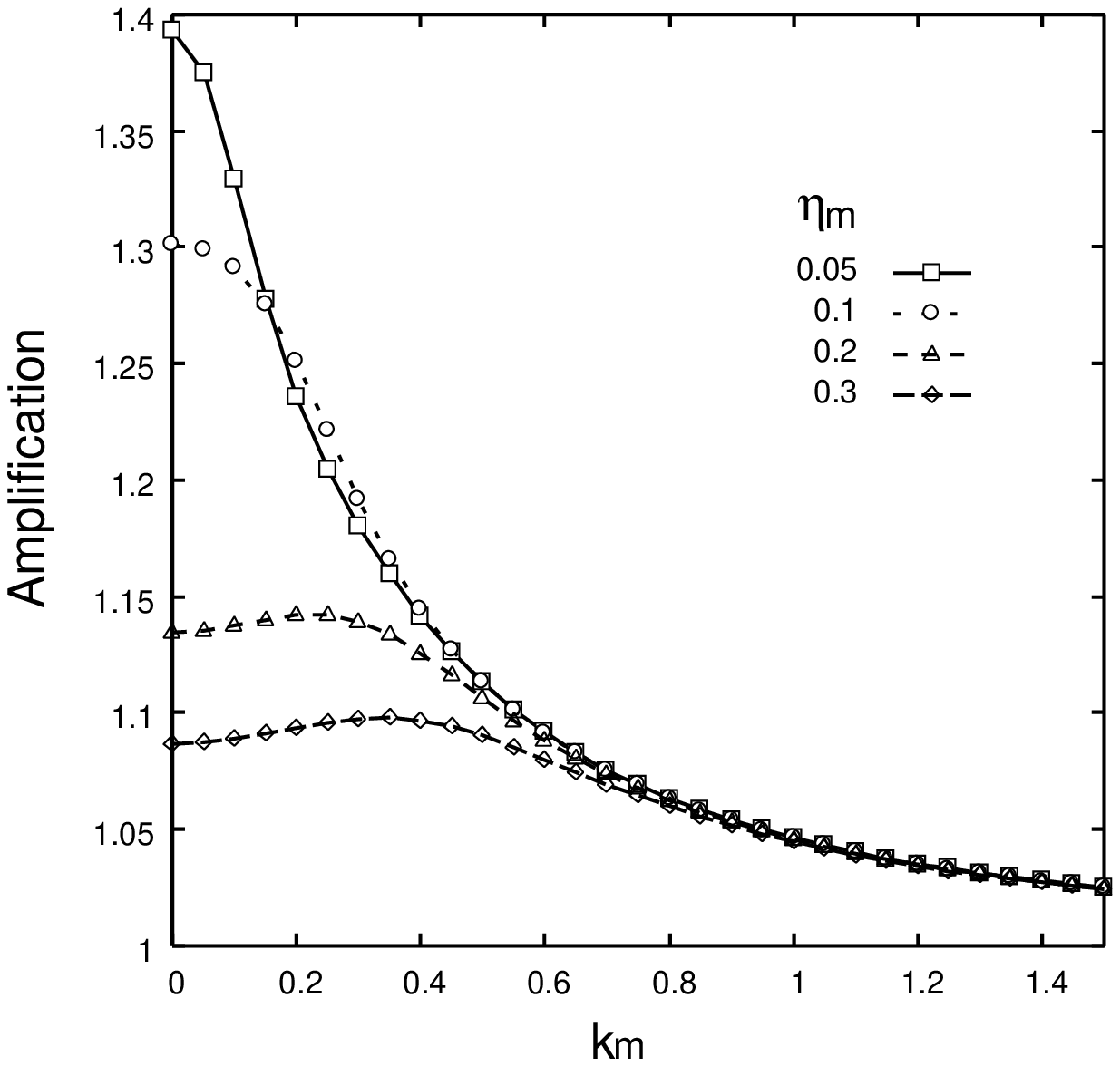}
\caption{Friction dependence of the amplification with $\lambda=1$ and $T_{m} = 0.1$.
The square designates data for $\eta_{m}=0.05$, the circle for $\eta_{m}=0.1$,
the triangle for $\eta_{m} = 0.2$ and the diamond for $\eta_{m} = 0.3$.
}
\label{fig:g=0:friction_dep}
	}
	\hfill
	\parbox{\halftext}{%
	\includegraphics[width=\halftext]{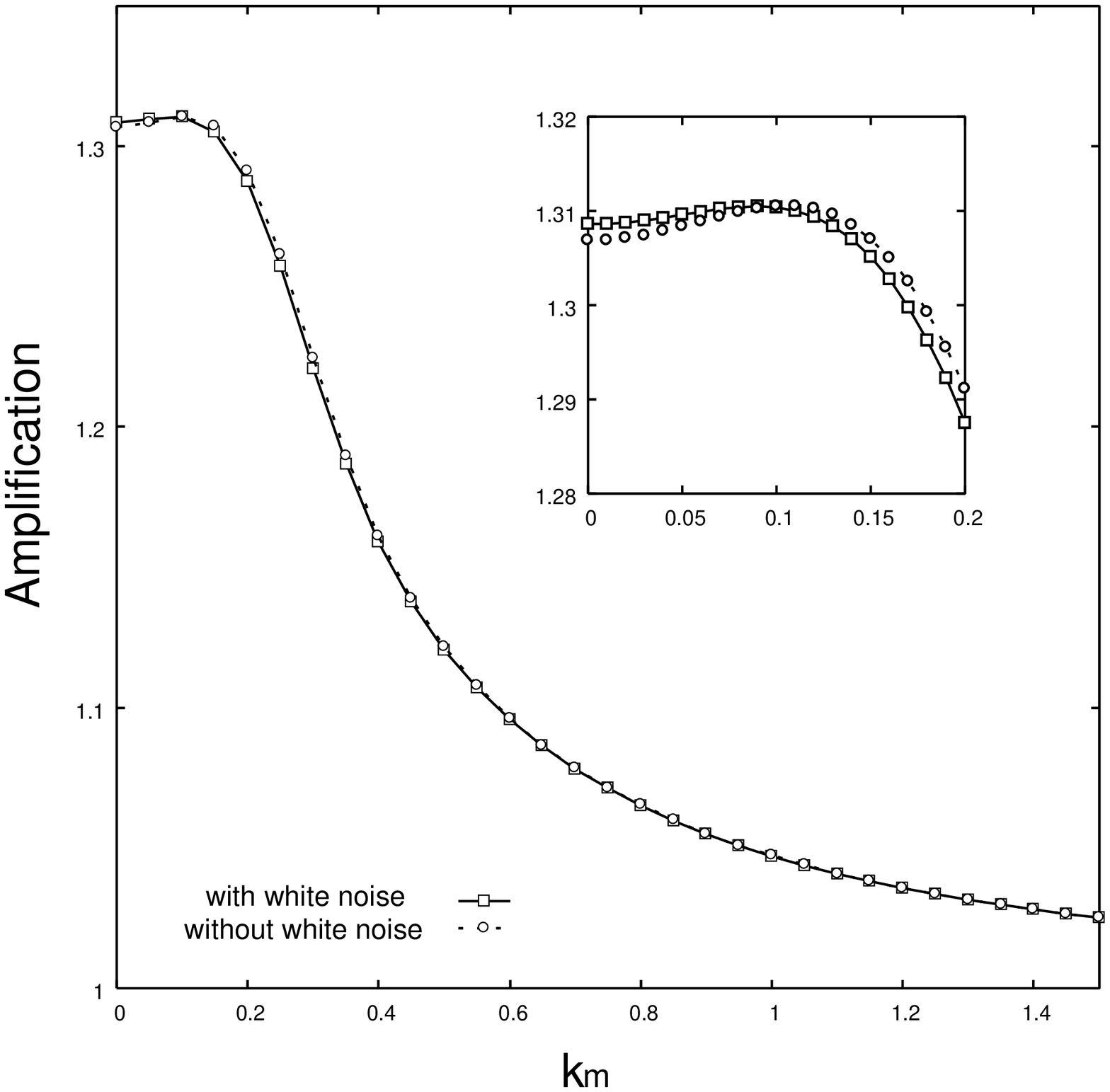}
\caption{
Comparison between the results with white noise and those without white noise.
The values of the parameters are $\lambda=0.1$ and  $T_{m}=0.1$.
The square designates data for 'with white noise' , the circle for 'without white noise'.
In the region of low $k_{m}$, 
the amplification with white noise is stronger than that without white noise.
}
\label{fig:g=0:precise}
	}
\end{figure}

Friction dependence is displayed in Fig. \ref{fig:g=0:friction_dep}.
Strong friction make $q_{2}^{T}$ small, because of the exponential factor in $q_{2}^{T}$.
Especially, the peak appears at $k_{m} \neq 0$ for strong friction, 
while it appears at $k_{m}=0$ for weak friction.
The suppression by friction is strong for the soft modes in this figure. 
This suppression is probably explained by the following fact.
The amplification depends not only on the initial magnitudes of $a^{T}$ and $q_{2}^{T}$, 
but also on the time dependences of these parameters, 
while the amplification  is determined by the (initial) magnitude of $a^{T}$ and $q_{2}^{T}$
in the Mathieu equation.

\begin{figure}
	\begin{center}
	\includegraphics[width=\textwidth]{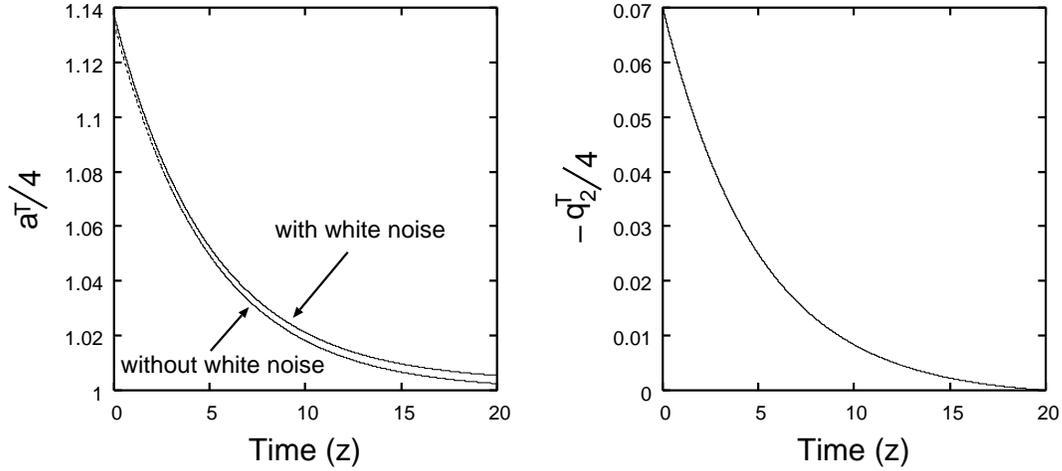}
	\end{center}
\caption{
Time evolution of $a^{T}/4$ and $-q^{T}_{2}/4$ with $k_{m}=0$, $\lambda=0.1$, $\eta_{m}=0.1$ and $T_{m}=0.1$. 
$a^{T}/4$ and $-q^{T}_{2}/4$ decrease quickly.
$-q^{T}_{2}/4$ approaches to zero asymptotically.
}
\label{fig:g=0:aq}
\end{figure}
It seems that white noise always suppresses the amplification. 
However, it enhances the amplification in some  exceptional cases. 
Figure \ref{fig:g=0:precise} displays the amplification with $\lambda=0.1$, $\eta_{m}= 0.1$ and $T_{m}=0.1$.
Thin curve shows the amplification with white noise and dotted curve without white noise.
In the region of $k_{m} \le 0.1 $, 
the amplification with white noise is slightly stronger than that without white noise.
This inversion seems the contradiction to the theory of the Mathieu equation, 
because $a^{T}$ with white noise is always larger than $a^{T}$ without white noise 
and because $a^{T}$ is always larger or equal to 4. 
In the $g=0$ cases, the amplification is determined by $a^{T}/4$ and $-q_{2}^{T}/4$. 
Figure \ref{fig:g=0:aq} displays the time dependence of $a^{T}/4$ and $-q_{2}^{T}/4$
 with $k_{m}=0$, $\lambda=0.1$, $\eta_{m}=0.1$ and $T_{m}=0.1$. 
It is easily found that $-q_{2}^{T}/4$ goes to zero quickly in a few oscillations of the condensate.
This fact implies that
the magnitude of the amplification is determined 
not only by the magnitudes of $a^{T}$ and $q_{2}^{T}$, but also by the time dependence of them.
Then, the inversion does not indicate the contradiction and happens in some cases.

\begin{figure}[htb]
	\parbox{\halftext}{%
	\includegraphics[width=\halftext]{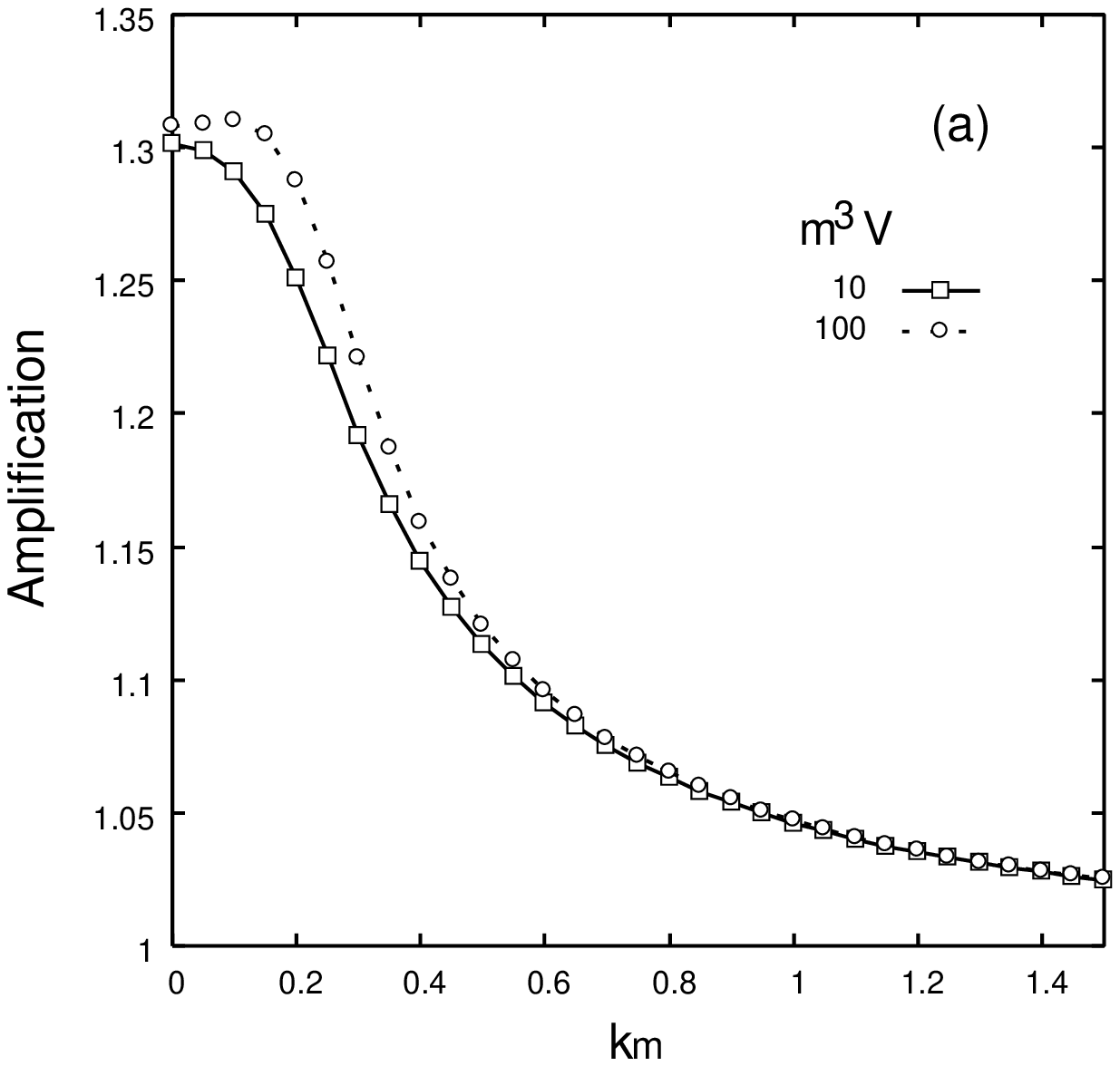}
	}
	\hfill
	\parbox{\halftext}{%
	\includegraphics[width=\halftext]{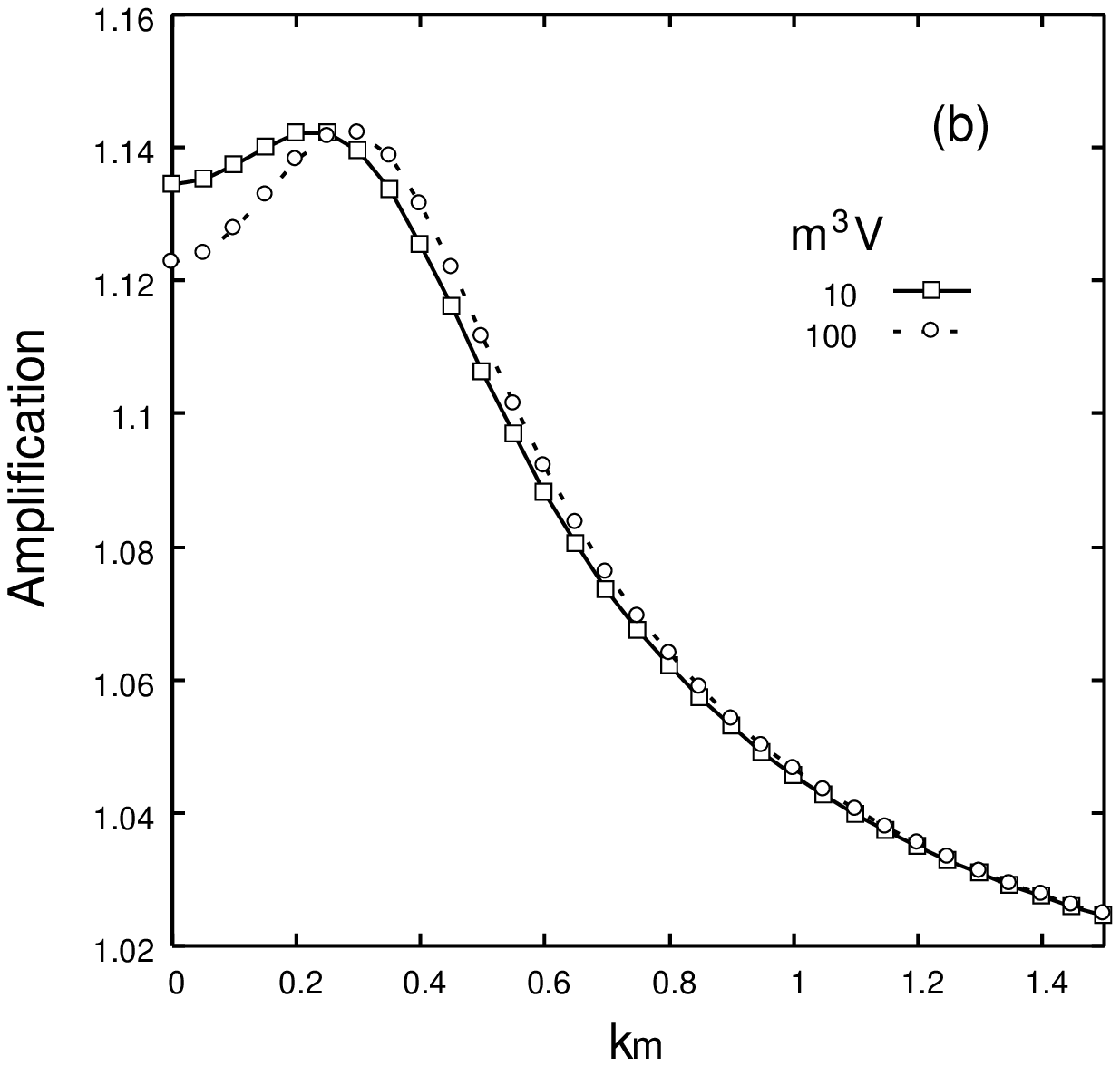}
	}
\caption{
Volume dependence of the amplification with $\lambda=1$ and $T_{m}=0.1$
for (a) $\eta_{m} = 0.1$ and (b) $\eta_{m} = 0.2$. 
The square designates data for $m^{3} V = 10$ and 
the circle for $m^{3} V = 100$.
}
\label{fig:Sym_m3V}
\end{figure}
Figure \ref{fig:Sym_m3V} displays the amplification for 
$m^{3}V=10$ and $100$ with $\lambda=1$ and $T_{m}=0.1$.
Figure \ref{fig:Sym_m3V} (a) displays the amplifications with $\eta_{m}=0.1$. 
Figure \ref{fig:Sym_m3V} (b) displays with $\eta_{m}=0.2$.
As shown in Fig. \ref{fig:Sym_m3V} (a), 
the amplification for $m^{3}V = 100$ is slightly stronger in wide range of $k_{m}$  
than that for $m^{3}V = 10$, because large volume reduces the magnitude of the white noise.
However, 
the magnitude of the amplification for  $m^{3}V = 100$ is smaller than that for $m^{3}V = 10$
in the region of small $k_{m}$ 
as shown in Fig. \ref{fig:Sym_m3V} (b).
As discussed in the previous paragraph, 
this inversion is probably caused by the quick decrease of $q_{2}^{T}$.

In any case, the fields are amplified in the soft modes.
The amplification decreases monotonically as a function of $k_{m}$ 
in the region of large $k_{m}$ and goes to 1 asymptotically as $k_{m}$ goes to infinity.

\begin{figure}[htb]
	\begin{center}
	\includegraphics[width=\halftext]{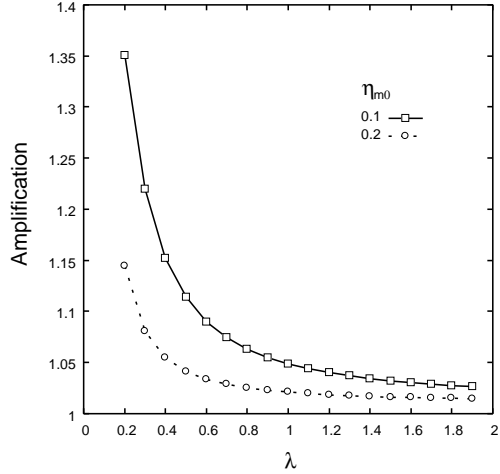}
	\end{center}
\caption{
Amplification in the cases that $\eta_{m}$ is proportional to $\lambda$.
The values of the parameters are $k_{m}=0$ and $T_{m}=0.1$. 
The friction is parametrized as $\eta_{m}=\eta_{m0} \lambda$.
The square designates data for $\eta_{m0}=0.1$ and the circle for $\eta_{m0}=0.2$. 
}
\label{fig:g=0:depend_on_lambda}
\end{figure}
$\eta_{m}$ in the above calculations is independent of $\lambda$. 
In the last part of this subsection, 
we investigate the amplification in the $\lambda$ dependent cases that 
$\eta_{m}$ is proportional to $\lambda$ as $\eta_{m} = \eta_{m0} \lambda$,
where $\eta_{m0}$ is independent of $\lambda$.
The $\lambda$ dependence of the amplification for $\eta_{m0}=0.1$ and $0.2$
are displayed in Fig. \ref{fig:g=0:depend_on_lambda}.
In Fig. \ref{fig:g=0:depend_on_lambda}, the amplification is strong in the weak coupling region
and weakens quickly with the increase of $\lambda$.
It is clear that 
strong friction causes this suppression,
because the friction decreases $q_{2}^{T}$.

\subsection{The $g_{m}=\frac{3 \sqrt{2 \lambda}}{2}$ cases}
\label{subsec:gm_ne_0}
In this subsection, we investigate the amplification in the $g_m = \frac{3 \sqrt{2 \lambda}}{2}$ cases.
$q_{1}^{T}$ is not zero in the present cases. [See eq. (\ref{eqn:q1T}).]  
The ratio $\left| q_{1}^{T}/q_{2}^{T} \right|$ is larger than 1 in almost the time,
especially, in the late time. 
Therefore, $q_{1}^{T}$ plays an important role on the amplifications of the fields .

\begin{figure}[htb]
	\parbox{\halftext}{%
	\includegraphics[width=\halftext]{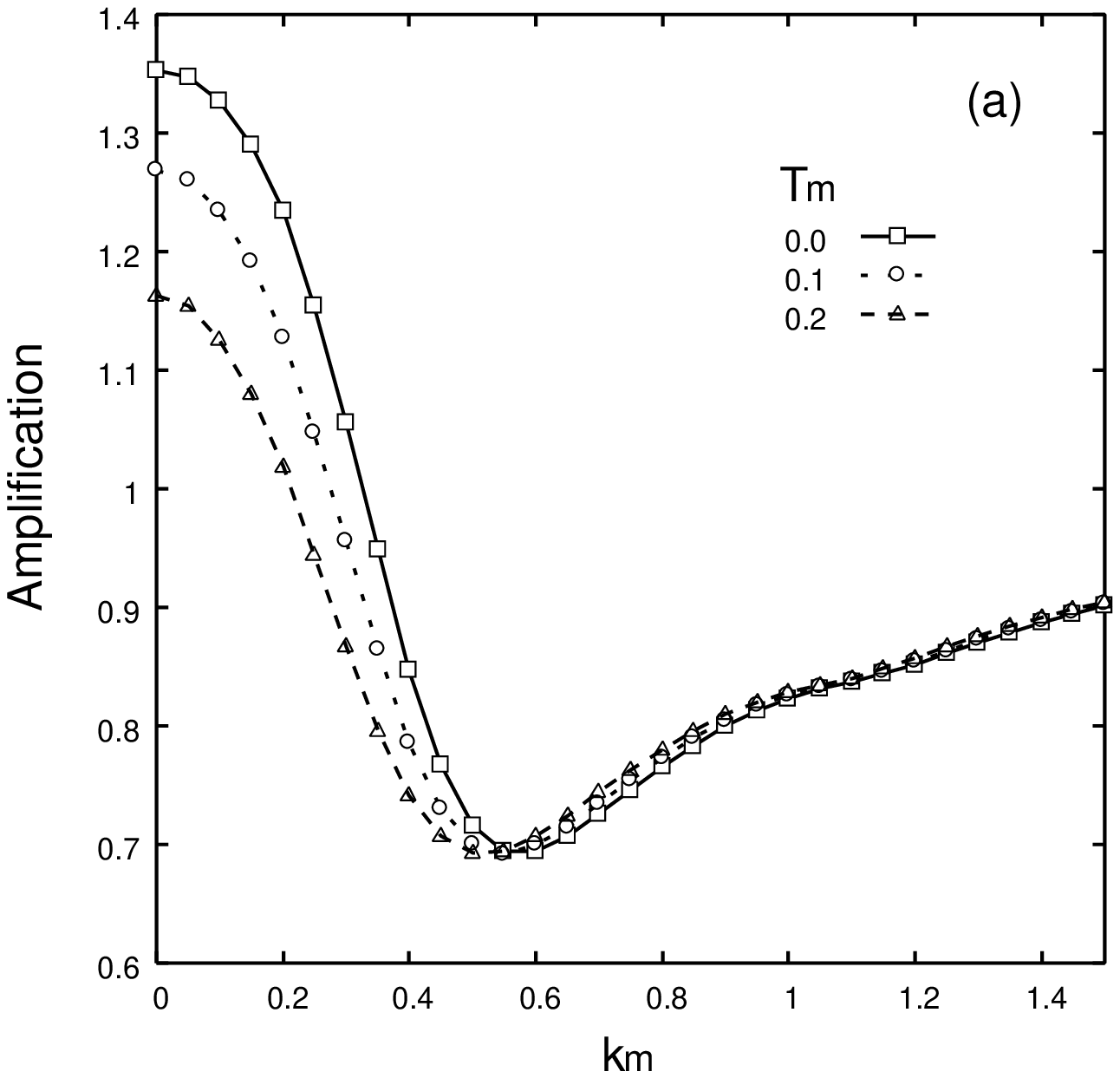}
	}
	\hfill
	\parbox{\halftext}{%
	\includegraphics[width=\halftext]{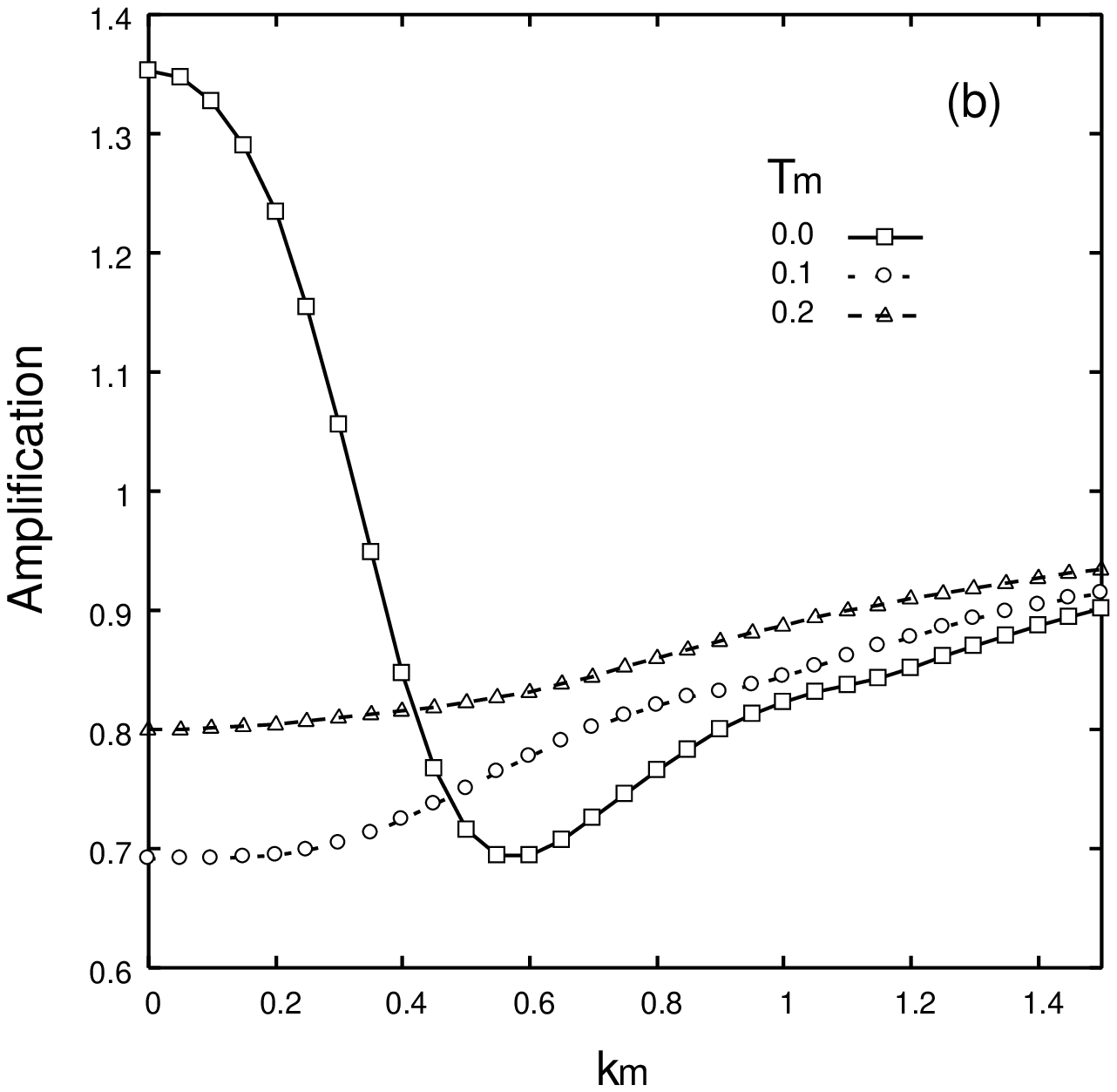}
	}
\caption{
Temperature dependence of the amplification with $\eta_{m}=0.2$. 
Figure \ref{fig:BR_TmdepL} (a) is depicted for $\lambda=1$ and Fig. \ref{fig:BR_TmdepL} (b) for $\lambda=10$.
The square designates data for $T_{m}=0$, the circle for $T_{m}=0.1$
and the triangle for $T_{m}=0.2$.
}
\label{fig:BR_TmdepL}
\end{figure}
Figure \ref{fig:BR_TmdepL} displays the temperature dependence of the amplification with $\eta_{m}=0.2$ 
for (a) $\lambda=1$ and (b) $\lambda=10$.
The amplification in the region of low $k_{m}$ is suppressed and   
this suppression is strong for large $\lambda$.
Therefore, the amplification may occur only for weak $\lambda$ at finite temperature. 
Dips are found around $k_{m} = 0.6$ at $T_{m}=0$ in the present cases,
while there is no dip in the $g_{m}=0$ cases.
Note that there is no effect of white noise at $T_{m}=0$. 
The curves accord well in the region of $k_{m} \ge 1$ for $\lambda=1$ 
and tend to 1 as $k_{m}$ increases.
This tendency in the region of large $k_{m}$ is also displayed for $\lambda = 10$.
This can be explained by the $k_{m}$ dependence of $a^{T}$ [eq. (\ref{eqn:numerical:aT})].
We remember that the Mathieu equation has the following form:
\begin{equation}
\partial_{z}^{2} y(z) + \left( a - 2q \cos(2z) \right) y(z) = 0.
\label{eqn:mathieu_eq}
\end{equation}
The stable region of the Mathieu equation spreads 
according as $a$ increases. 
Therefore, the solution of the Mathieu equation is similar to a sine function for large $a$, 
because the cosine term in the Mathieu equation is negligible.
The amplitude of the field is equal to the initial value 1, 
because no amplification occurs for large $a$ in the Mathieu equation.  
The time dependence of the solution of eq. (\ref{eq:numerical}) for large $k_{m}$ is explained on similar lines,
because eq. (\ref{eq:numerical}) is similar to eq. (\ref{eqn:mathieu_eq}).
\begin{figure}[htb]
	\parbox{0.33\textwidth}{%
	\includegraphics[width=0.33\textwidth]{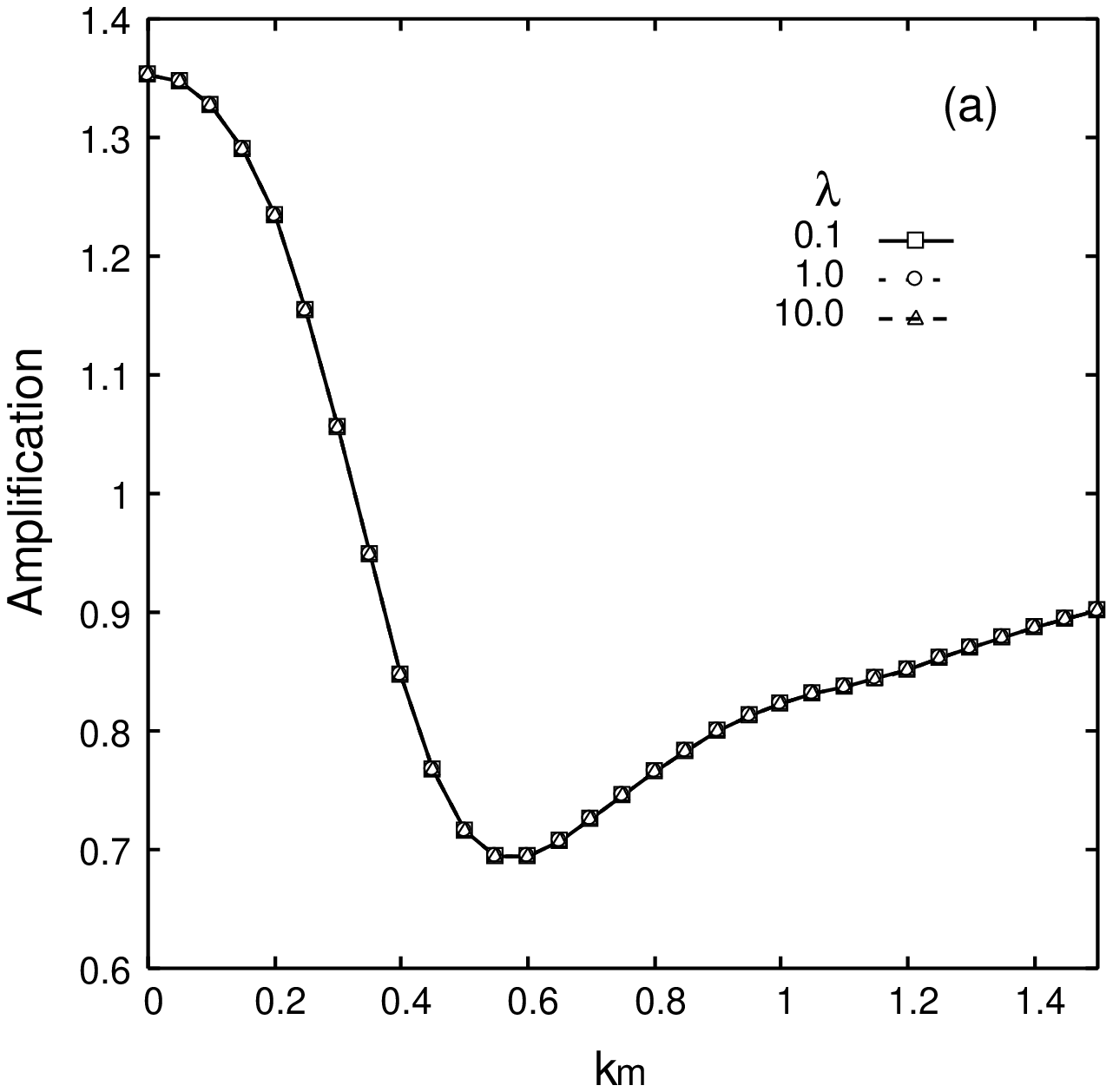}
	}
	\parbox{0.33\textwidth}{%
	\includegraphics[width=0.33\textwidth]{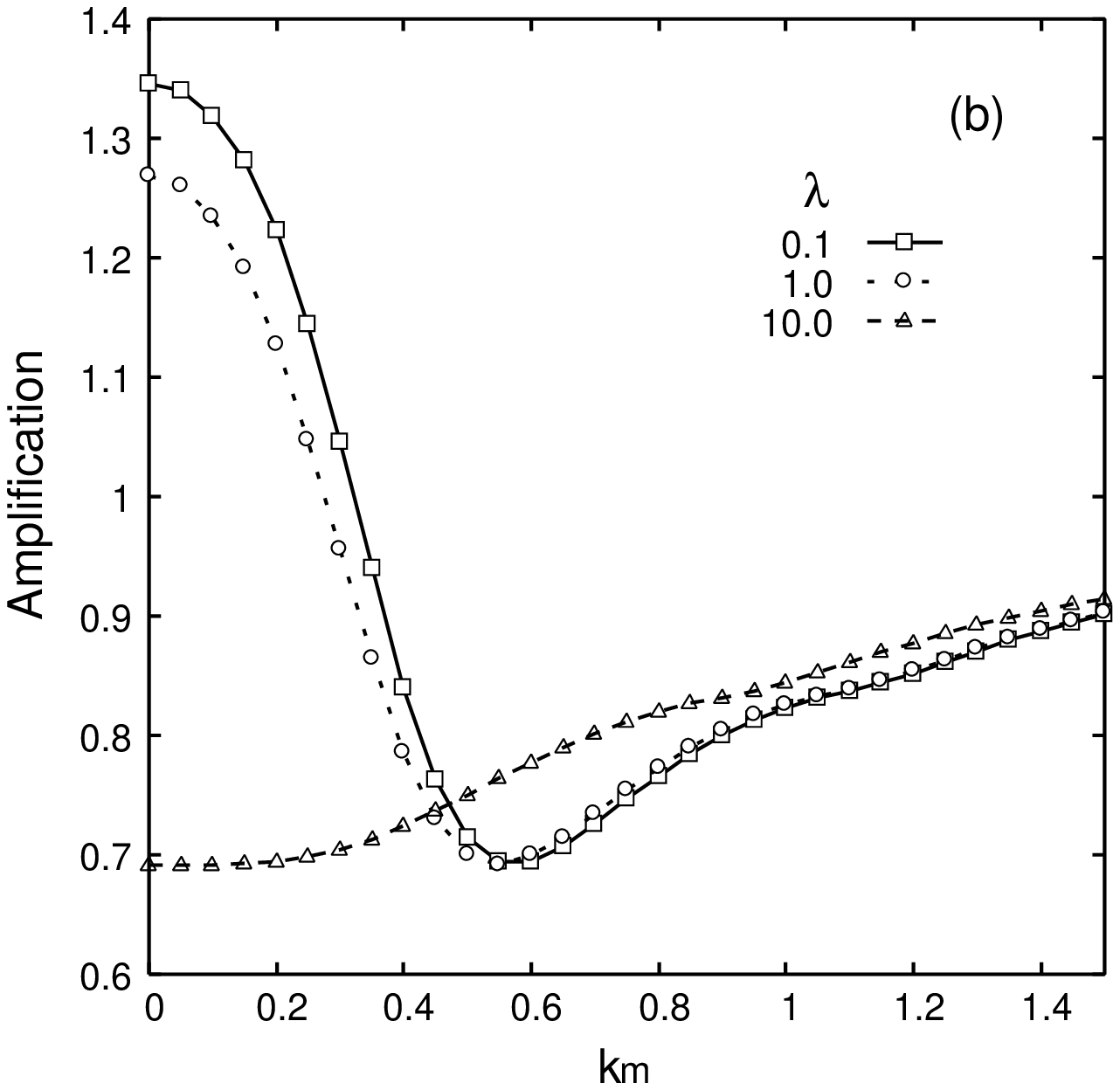}
	}
	\parbox{0.33\textwidth}{%
	\includegraphics[width=0.33\textwidth]{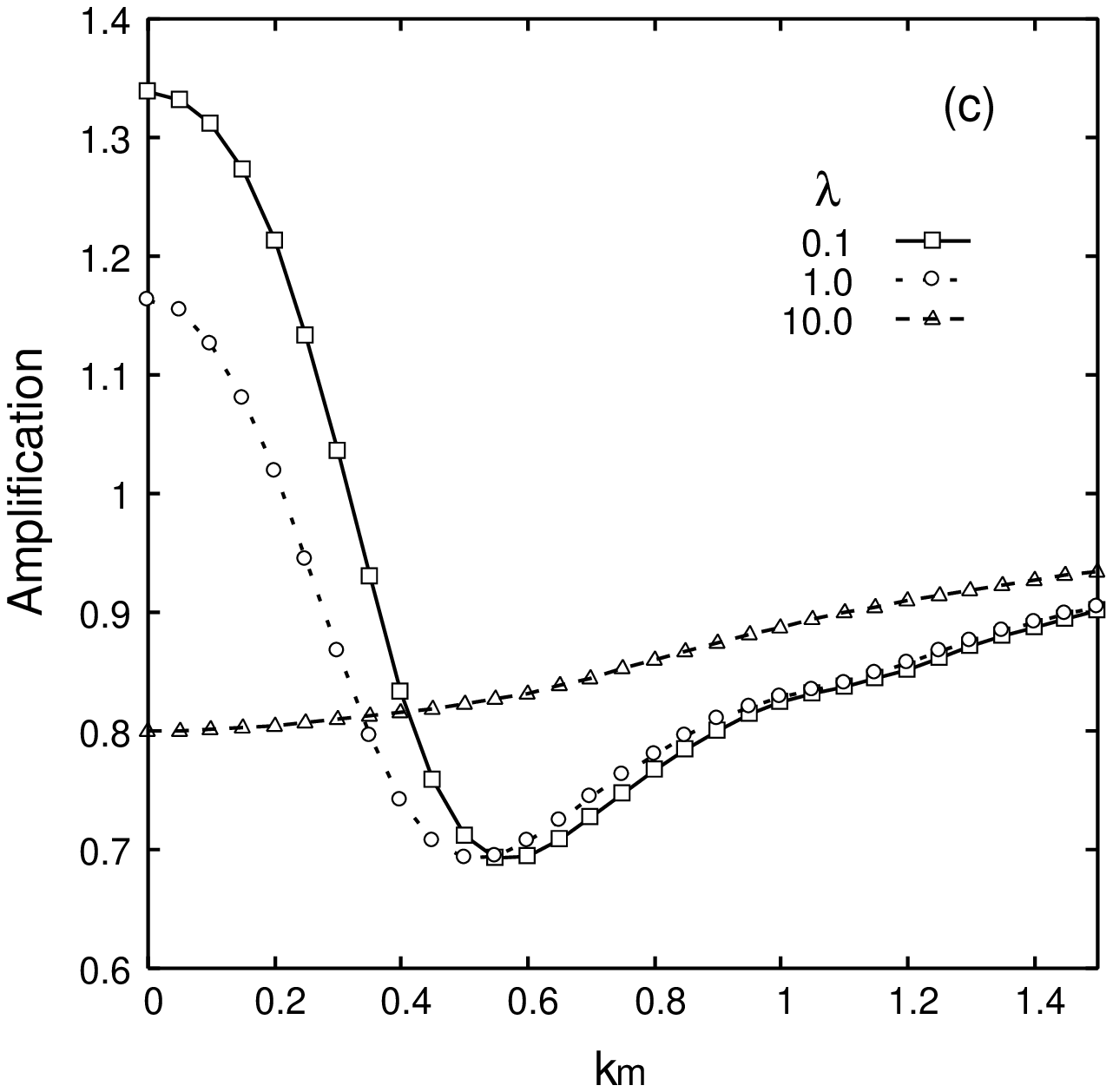}
	}
\caption{
Coupling dependence of the amplification with $\eta_{m}=0.2$. 
Figure \ref{fig:BR_RdepTm} (a) is depicted for $T_{m}=0$, Fig. \ref{fig:BR_RdepTm} (b) for $T_{m}=0.1$ 
and Fig. \ref{fig:BR_RdepTm} (c) for $T_{m}=0.2$.
The square designates data for $\lambda=0.1$, the circle for $\lambda=1$ 
and the triangle for $\lambda=10$.
}
\label{fig:BR_RdepTm}
\end{figure}
In Fig. \ref{fig:BR_TmdepL}, 
the amplification in the region of low $k_{m}$ for $T_{m} \neq 0$ is weaker than that for $T_{m}=0$.
This behavior is a nontrivial result from the equation similar to the Mathieu equation.
The amplification for $\lambda=10$  does not decrease monotonically as the temperature increases.
This reduction is most likely explained by the oscillation of the fields and 
the time dependence of the parameters,  $a^{T}$, $q_{1}^{T}$ and $q_{2}^{T}$.

\begin{figure}[htb]
	\parbox{\halftext}{%
	\includegraphics[width=\halftext]{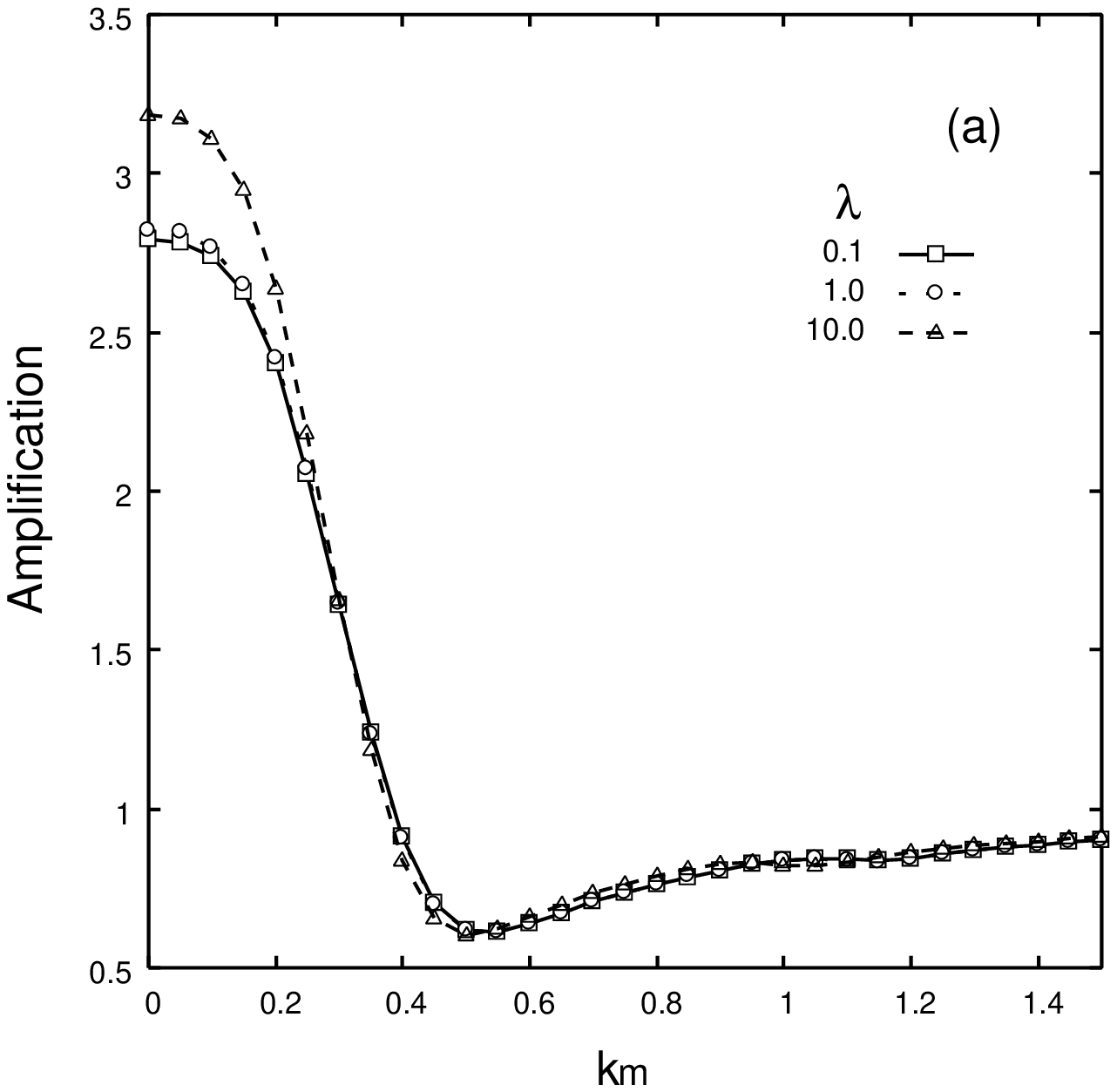}
	}
	\hfill
	\parbox{\halftext}{%
	\includegraphics[width=\halftext]{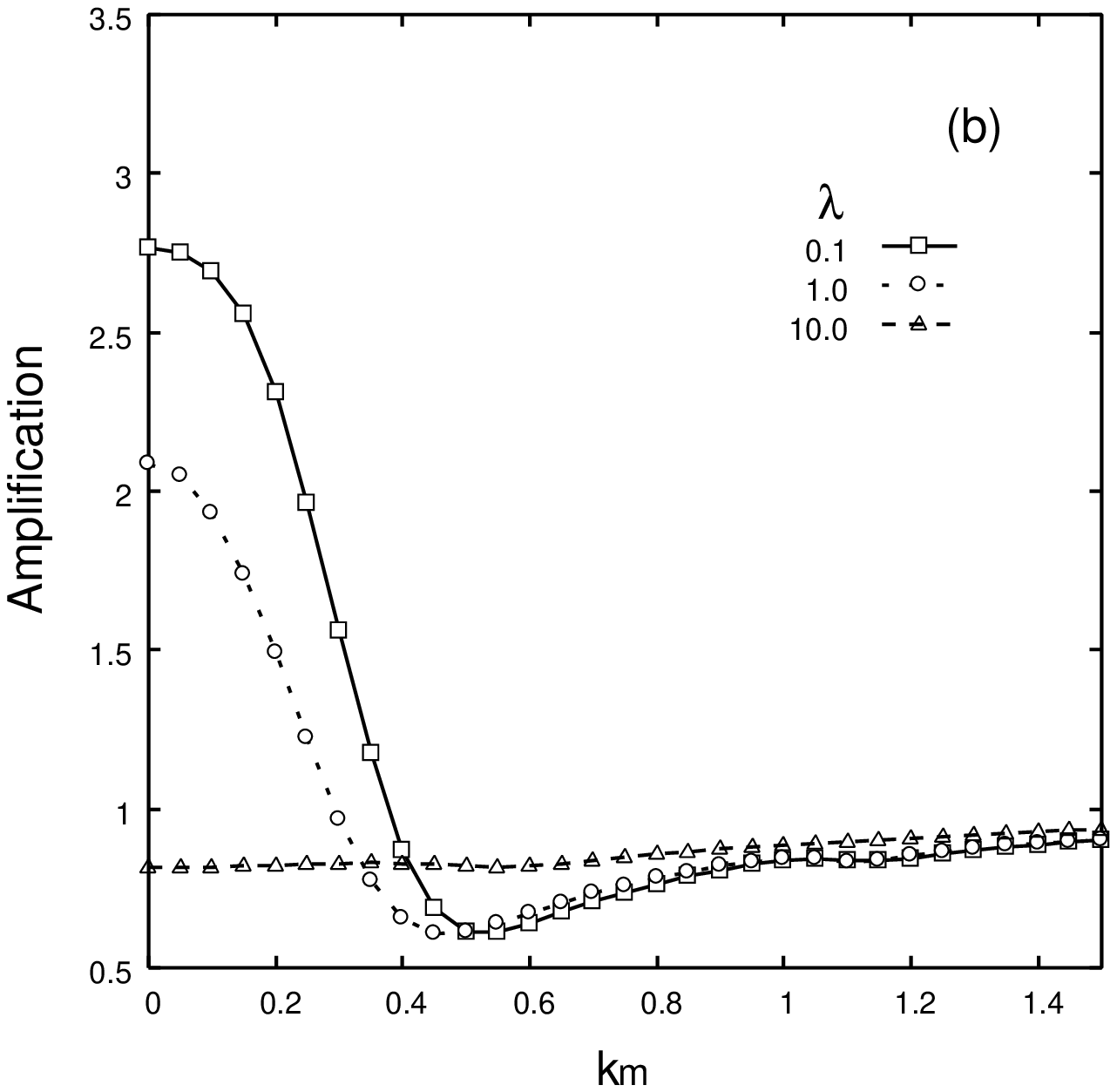}
	}
\caption{
Comparison between 
(a) the coupling dependence of the amplification without white noise and (b) that with white noise.
The values of the parameters are $T_{m}=0.2$ and $\eta_{m}=0.1$.
The square designates data for $\lambda=0.1$, the circle for $\lambda=1$ 
and the triangle for $\lambda=10$.
}
\label{fig:BR_RdepEtam}
\end{figure}
The amplification for various values of $\lambda$ is displayed in Fig. \ref{fig:BR_RdepTm}
with $\eta_{m}=0.2$ and various values of $T_{m}$.
The amplification at $T_{m}=0$ is displayed in Fig. \ref{fig:BR_RdepTm}(a).
The curves in this figure are the same, because eq. (\ref{eqn:numerical:aT}) is $\lambda$ independent
at $T_{m}=0$.
Indeed, 
$\lambda C_{m}^{2}$ is $\lambda$ independent, 
because the initial amplitude of the condensate $C_{m}$ is proportional to $\lambda^{-1/2}$.
Moreover, the coupling $g$ and the mass squared $m^{2}$ are 
$3 \lambda v$ and $2 \lambda v^{2}$, respectively. 
Thus, $\Phi_{0}$ is $\lambda$ independent,
because $\Phi_{0}$ satisfies the equation $\Phi_{0}^{3} + 3 v \Phi_{0}^{2} + 2 v^{2} \Phi_{0} = 0$.
Therefore, $a^{T}$, $q_{1}^{T}$ and $q_{2}^{T}$ are $\lambda$ independent at $T_{m}=0$.
We also remember that $\eta_{m}$ is $\lambda$ independent. 
(This value is given by hand in the present numerical calculations.)
From this consideration, we find that 
eq. (\ref{eq:numerical}) at $T_{m}=0$ is independent of $\lambda$.
(This coincidence is artificial in the sense that the result depends on 
the initial amplitude of the condensate and the $\lambda$ dependence of friction.)
The $\lambda$ dependences at $T_{m}=0.1$ and  $T_{m}=0.2$ are displayed 
in Fig. \ref{fig:BR_RdepTm}(b) and Fig. \ref{fig:BR_RdepTm}(c) respectively.
The amplification weakens in the region of low $k_{m}$ according as $\lambda$ increases.
Especially, the amplification is strongly suppressed for $\lambda=10$.

Figure \ref{fig:BR_RdepEtam} displays the amplification for various values of $\lambda$
with $\eta_{m}=0.1$ and $T_{m}=0.2$.
Figure \ref{fig:BR_RdepEtam}(a) displays the amplification without white noise and 
Fig. \ref{fig:BR_RdepEtam}(b) with white noise.
The strength of the amplification increases with the coupling $\lambda$ in Fig. \ref{fig:BR_RdepEtam}(a), 
while that decreases with the coupling $\lambda$ in Fig. \ref{fig:BR_RdepEtam}(b).
The $\lambda$ dependence  of the amplification is opposite. 
This result implies that white noise can change the growth of the fields.
(Therefore the particle distribution is probably different.)

\begin{figure}[htb]
	\parbox{\halftext}{%
	\includegraphics[width=\halftext]{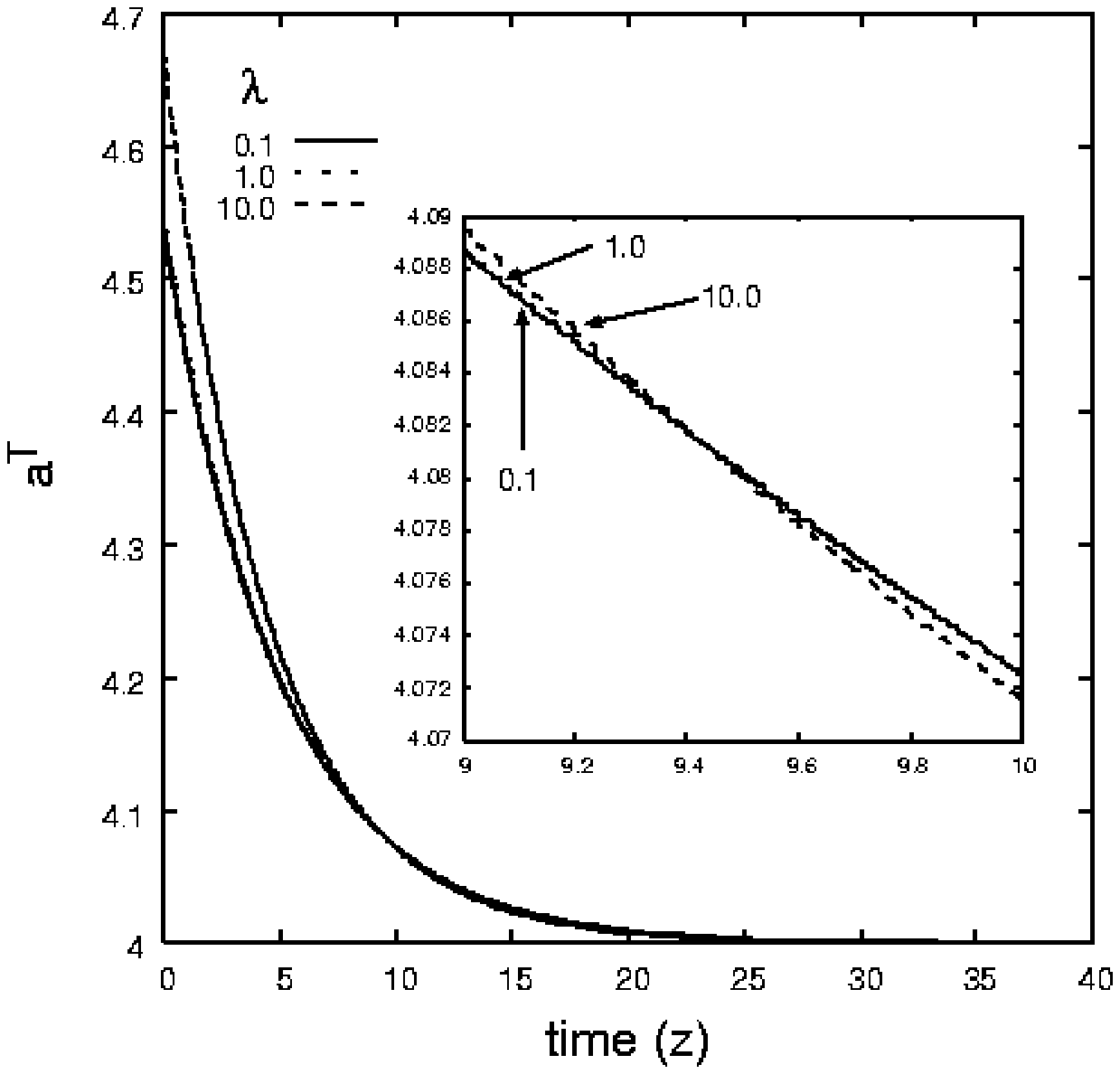}
\caption{
Time evolution of $a^{T}$ with $T_{m}=0.2$ and $\eta_{m}=0.1$ 
for various values of $\lambda$.
These curves intersect at different points.
}
\label{fig:timedep_of_aT}
	}
	\hfill
	\parbox{\halftext}{%
	\includegraphics[width=\halftext]{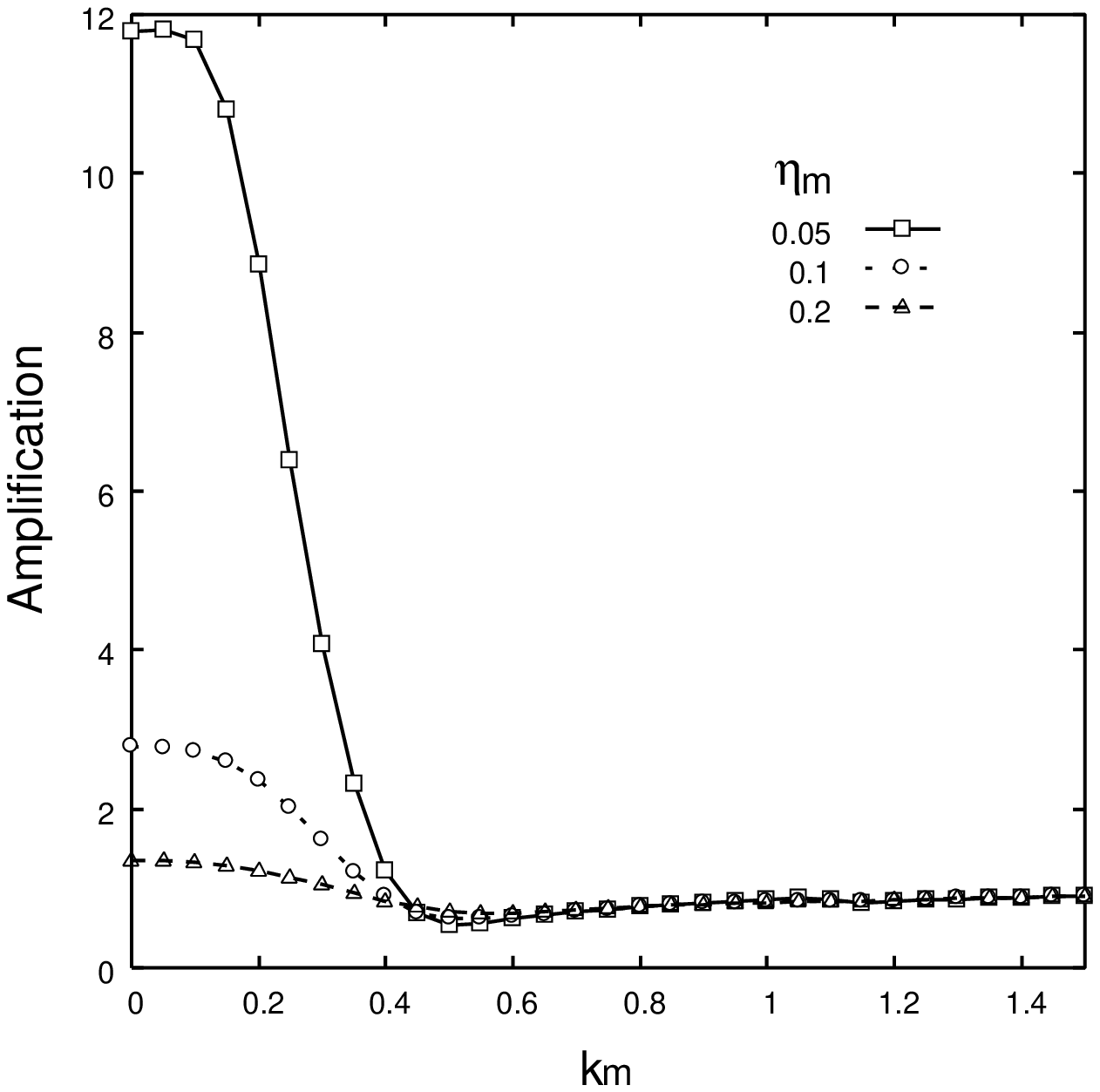}
\caption{
Friction dependence of the amplification with $\lambda=0.1$ and $T_{m}=0.1$. 
The square designates data for $\eta_{m}=0.05$, the circle for $\eta_{m}=0.1$
and the triangle for $\eta_{m}=0.2$.
}
\label{fig:BR_Etam_dep}
	}
\end{figure}
Figure \ref{fig:timedep_of_aT} is depicted to present the time dependence of 
$a^{T}$ without white noise for various values of $\lambda$ with $\eta_{m}=0.1$ and $T_{m}=0.2$. 
$a^{T}$ for $\lambda=10$ is largest in the early time, but it is smallest in the late time. 
(Note that the curves in Fig. \ref{fig:timedep_of_aT} intersect at different points.)
The $z$ dependence of the parameters implies that 
the field can grow even for large $\lambda$ as shown in Fig. \ref{fig:BR_RdepEtam}(a).

Figure \ref{fig:BR_Etam_dep} displays the amplification for some $\eta_{m}$ with $\lambda=0.1$ and $T_{m}=0.1$.
The strength of the amplification decreases with the friction $\eta_{m}$.
Apparently, the friction makes the amplification weak. 
The reason is the same as discussed in the previous subsection.

\begin{figure}[htb]
	\parbox{\halftext}{%
	\includegraphics[width=\halftext]{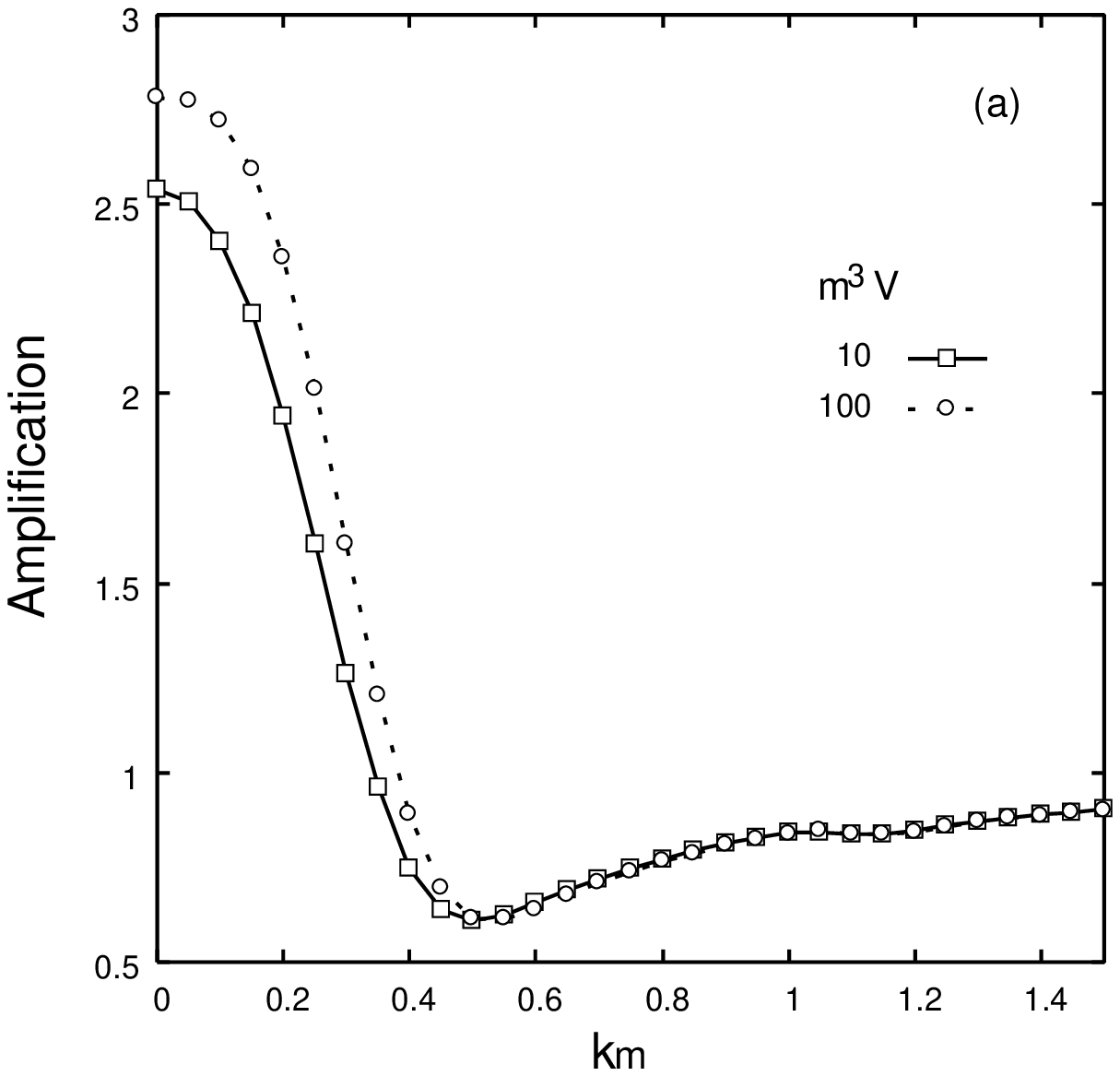}
	}
	\hfill
	\parbox{\halftext}{%
	\includegraphics[width=\halftext]{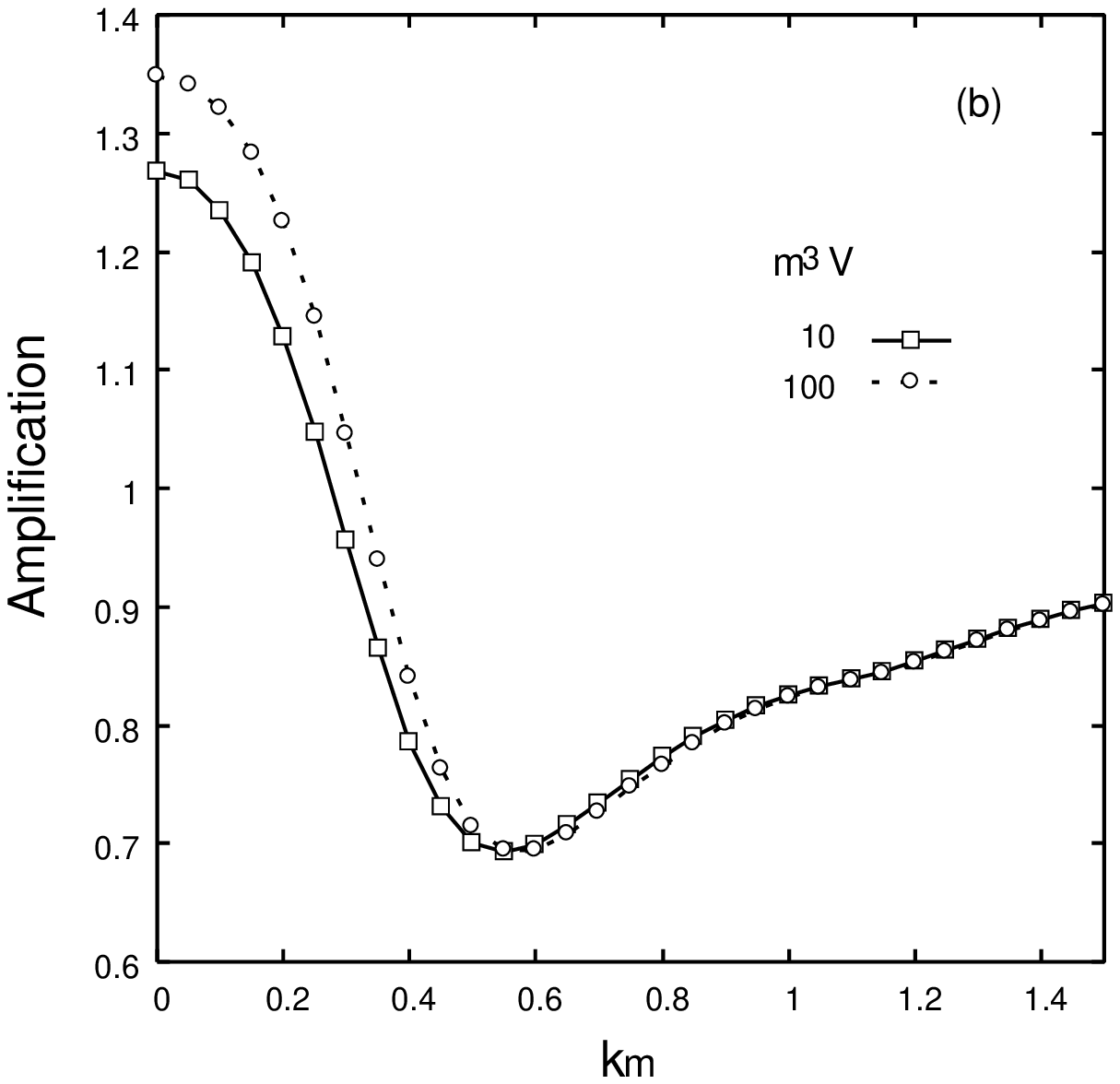}
	}
\caption{
Volume dependence of the amplification with $\lambda=1$ and $T_{m}=0.1$
for (a) $\eta_{m} = 0.1$ and (b) $\eta_{m} = 0.2$.
The square designates data for $m^{3} V = 10$ and 
the circle for $m^{3} V = 100$.
}
\label{fig:BR_m3V}
\end{figure}
Figure \ref{fig:BR_m3V} displays the amplification for different volumes with
 (a) $\eta_{m}=0.1$ and (b) $\eta_{m}=0.2$.
The parameters are $\lambda=1$ and  $T_{m}=0.1$.
In both cases, the amplification for $m^{3}V=10$ is weaker than that for $m^{3}V=100$.
(The amplification for $m^{3}V=10$ is slightly stronger than that for $m^{3}V=100$ around $k_{m} =0.7$.)
$a^{T}$ for $m^{3}V=100$ is smaller than that for $m^{3}V=10$, 
because the magnitude of white noise is proportional to $V^{-1/2}$ as in eq. (\ref{eqn:randomforce_corr}).
Therefore, the fields are amplified strongly for large $V$.

\begin{figure}[htb]
	\begin{center}
	\includegraphics[width=\halftext]{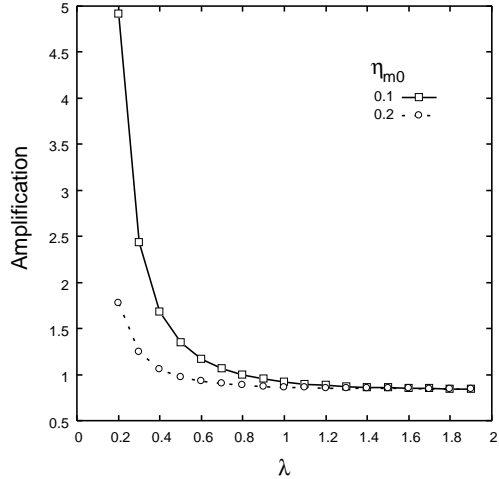}
	\end{center}
\caption{
Amplification in the cases that $\eta_{m}$ is proportional to $\lambda$.
The values of the parameters are $k_{m}=0$ and $T_{m}=0.1$.
The friction is parametrized as  $\eta_{m}=\eta_{m0}\lambda$.
The square designates data for $\eta_{m0}=0.1$ and  the circle for $\eta_{m0}=0.2$.
}
\label{fig:Lambda_dep}
\end{figure}
Finally, as shown in Fig. \ref{fig:BR_Etam_dep},
the amplification for various values of $\lambda$ is displayed in Fig. \ref{fig:Lambda_dep} with $k_{m}=0$ and $T_{m}=0.1$
in the case that the magnitude of the friction is proportional to $\lambda$. 
The friction is parametrized as $\eta_{m} = \eta_{m0} \lambda$ as in Sec. \ref{subsec:gm=0}, 
while the friction is $\lambda$ independent in the previous calculations in this subsection.  
The $\lambda$ dependence of the amplification in the $g_{m} = 3 \sqrt{2\lambda} / 2$ cases is similar to 
that in the $g_{m} = 0$ cases.
The amplification depends on the magnitude of the friction strongly,
because the friction affects the coefficients, $a^{T}$, $q_{1}^{T}$ and $q_{2}^{T}$ 
through the exponential factors.
Therefore, 
the strength of the amplification decreases with $\lambda$ as shown in  Fig. \ref{fig:Lambda_dep}.

\section{Conclusions}
\label{sec:conclusions}
We investigated the effects of white noise on parametric resonance in $\lambda \phi^{4}$ theory.
The potential $V(\phi)$ in this study is 
$\frac{1}{2} m^{2} \phi^{2} + \frac{1}{3} g \phi^{3} + \frac{1}{4} \lambda \phi^{4}$.
The effects of white noise for the fluctuation field $\psi$ defined in eq. (\ref{eqn:divide}) 
are included through the motion of the condensate.
White noise modifies the mass term in the equation of motion for the soft modes. 

It was found from the present numerical study
that the amplitudes of the fields for the soft modes are amplified due to parametric resonance
in both the $g = 0$ and $g = \frac{3\sqrt{2\lambda}}{2} m$ cases. 
The white noise  suppresses the amplification in almost all the cases.
However, in some cases, the amplification with white noise is slightly stronger than that without white noise.
The strength of the amplification decreases with the temperature, the friction and the coupling $\lambda$.
This result lead to an opposite conclusion:
white noise suppresses the amplification in almost all the cases in $\lambda \phi^{4}$ theory,
while white noise enhances in some other theories.
The suppression is caused by the $\Phi^{2} \phi_{s}$ term
in the equation of motion for the soft modes in $\lambda \phi^{4}$ theory. 
Therefore, the form of the coupling between the condensate and finite modes is essential to the amplification
due to parametric resonance with noise.

In the present numerical study, 
we found two different points between the results in the $g=0$ cases and 
those in the $g = \frac{3\sqrt{2\lambda}}{2} m $ cases:
1) The fields are always amplified in the $g=0$ cases. 
Contrarily, the fields are suppressed around $k_{m} = 0.6$ in the $g = \frac{3\sqrt{2\lambda}}{2} m$ cases.
Thus, the relative ratio of the amplitudes for different modes is emphasized in the latter cases.
In other words, the momentum dependence of the field amplitudes is emphasized.
2) The amplification of the field is maximal at $k_{m} \neq 0$  in some $g=0$ cases,
while that is maximal at $k_{m} \sim 0$ in $g = \frac{3\sqrt{2\lambda}}{2} m$ cases
when the amplification occurs.
It is possible to distinguish by these differences
whether the system is on the $g=0$ state or not. 

In this paper, the effects of the linear term of $\xi$ are not taken into account, 
because the noise is averaged.
Back reaction and effects of the nonlinear terms are also ignored. 
We would like to include these effects in the future studies. 


\end{document}